\documentclass[sn-mathphys,iicol,Numbered]{sn-jnl}

\usepackage{graphicx}%
\usepackage{multirow}%
\usepackage{amsmath,amssymb,amsfonts}%
\usepackage{amsthm}%
\usepackage{mathrsfs}%
\usepackage[title]{appendix}%
\usepackage{xcolor}%
\usepackage{textcomp}%
\usepackage{manyfoot}%
\usepackage{booktabs}%
\usepackage{algorithm}%
\usepackage{algorithmicx}%
\usepackage{algpseudocode}%
\usepackage{listings}%

\usepackage{savesym}
\savesymbol{tablenum}
\usepackage{siunitx}
\restoresymbol{SIX}{tablenum}

\usepackage{bm}

\begin{document}

\title[On the elastoplastic behavior in collisional compression of spherical dust aggregates]{On the elastoplastic behavior in collisional compression of spherical dust aggregates}


\author*[1]{\fnm{Sota} \sur{Arakawa}}\email{arakawas@jamstec.go.jp}

\author[2]{\fnm{Hidekazu} \sur{Tanaka}}

\author[3]{\fnm{Eiichiro} \sur{Kokubo}}

\author[4]{\fnm{Satoshi} \sur{Okuzumi}}

\author[5]{\fnm{Misako} \sur{Tatsuuma}}

\author[1]{\fnm{Daisuke} \sur{Nishiura}}

\author[1]{\fnm{Mikito} \sur{Furuichi}}

\affil*[1]{\orgdiv{Center for Mathematical Science and Advanced Technology}, \orgname{Japan Agency for Marine-Earth Science and Technology}, \orgaddress{\street{3173-25 Showa-machi, Kanazawa-ku}, \city{Yokohama}, \postcode{236-0001}, \country{Japan}}}

\affil[2]{\orgdiv{Astronomical Institute, Graduate School of Science}, \orgname{Tohoku University}, \orgaddress{\street{6-3 Aramaki, Aoba-ku}, \city{Sendai}, \postcode{980-8578}, \country{Japan}}}

\affil[3]{\orgdiv{Center for Computational Astrophysics}, \orgname{National Astronomical Observatory of Japan}, \orgaddress{\street{2-21-1 Osawa, Mitaka}, \city{Tokyo}, \postcode{181-8588}, \country{Japan}}}

\affil[4]{\orgdiv{Department of Earth and Planetary Sciences}, \orgname{Tokyo Institute of Technology}, \orgaddress{\street{2-12-1 Ookayama, Meguro}, \city{Tokyo}, \postcode{152-8550}, \country{Japan}}}

\affil[5]{\orgdiv{Interdisciplinary Theoretical and Mathematical Sciences Program}, \orgname{RIKEN}, \orgaddress{\street{2-1 Hirosawa, Wako}, \city{Saitama}, \postcode{351-0198}, \country{Japan}}}


\abstract{

Aggregates consisting of submicron-sized cohesive dust grains are ubiquitous, and understanding the collisional behavior of dust aggregates is essential.
It is known that low-speed collisions of dust aggregates result in either sticking or bouncing, and local and permanent compaction occurs near the contact area upon collision.
In this study, we perform numerical simulations of collisions between two aggregates and investigate their compressive behavior.
We find that the maximum compression length is proportional to the radius of aggregates and increases with the collision velocity.
We also reveal that a theoretical model of contact between two elastoplastic spheres successfully reproduces the size- and velocity-dependence of the maximum compression length observed in our numerical simulations.
Our findings on the plastic deformation of aggregates during collisional compression provide a clue to understanding the collisional growth process of aggregates.

}




\maketitle

\clearpage

\section{Introduction}

Aggregates consisting of micron- and submicron-sized cohesive dust grains are ubiquitous on Earth \cite{BROWN201265} and in space \cite{2016Natur.537...73B}.
Astronomical observations at near-infrared wavelengths have revealed that aggregates in planet-forming disks surrounding young stars are indeed composed of submicron-sized grains \cite{2022A&A...663A..57T, 2023ApJ...944L..43T}.
They grow through multiple mutual collisions; therefore, understanding the collisional behavior of dust aggregates is essential.
A considerable number of studies have been conducted using both laboratory experiments \cite{antonyuk2006impact, 2008ARA&A..46...21B, 2008ApJ...675..764L, antonyuk2010energy, 2011ApJ...736...34B, 2012Icar..221..310S, 2012Icar..218..688W, 2018ApJ...853...74S, 2022MNRAS.509.5641S} and numerical simulations \cite{1997ApJ...480..647D, 2009A&A...507.1023P, 2013A&A...559A..62W, 2021ApJ...915...22H, 2021PhRvE.103d2902V, chen2022simulations, OSINSKY2022127785, 2023A&A...670L..21A, 2023MNRAS.526..523B}.
Collisional outcomes of aggregates are classified into several categories, including sticking, bouncing, erosion, and fragmentation \cite{2010A&A...513A..56G}.

Both numerical simulations \cite{2011ApJ...737...36W, 2013A&A...551A..65S} and laboratory experiments \cite{2012Icar..218..688W, 2022MNRAS.509.5641S} have consistently reported that low-speed collisions of equal-mass aggregates result in either sticking or bouncing.
The outcomes of aggregate--aggregate collisions are stochastic, and several studies have investigated the size- and velocity-dependent sticking probability \cite{2008ApJ...675..764L, 2012Icar..218..688W, 2023ApJ...951L..16A}.
For instance, Arakawa et al.~\cite{2023ApJ...951L..16A} reported that the sticking probability decreases with increasing the aggregate size.
Consequently, bouncing could potentially limit the mass growth of aggregates through mutual collisions.

In addition to collision experiments and simulations, researchers have also delved into the quasistatic mechanical properties of bulk aggregates.
Compressive and tensile strengths of aggregates have been measured by laboratory experiments \cite{2004PhRvL..93k5503B, 2009ApJ...701..130G, 2009JGRE..114.9004Y, 2017P&SS..149...14O, 2018ApJ...860..123O, 2022MNRAS.512.3754F}, and elastic properties have been evaluated based on sound-speed measurements \cite{2012A&A...544A.138M}.
Furthermore, numerical simulations have been employed to investigate mechanical strengths \cite{2012A&A...541A..59S, 2013A&A...559A..19S, 2019ApJ...874..159T, 2023ApJ...953....6T}.
These mechanical properties are requisites when utilizing continuum models to simulate collisions of macroscopic aggregates consisting of billions of particles \cite{2004Icar..167..431S, 2010A&A...513A..58G}.

Bouncing collisions among porous aggregates result in significant energy loss \cite{1993Icar..106..151B}.
Local and permanent compaction occurs near the contact area upon collision \cite{2009ApJ...696.2036W, TOYODA2024115964}.
Therefore, understanding the bouncing behavior entails exploring the plastic deformation of aggregates during collisional compression.

In this study, we perform numerical simulations of collisions between two aggregates and investigate their compressive behavior.
We reveal that the maximum compression length is proportional to the radius of aggregates and increases with the collision velocity.
The size- and velocity-dependence of the maximum compression length is compared with a theoretical model of contact between two elastoplastic spheres \cite{doi:10.1080/14786443008565033}, and we succeed in reproducing the dependence observed in our numerical simulations.

\section{Discrete Element Method}

We perform three-dimensional numerical simulations of collisions between two dust aggregates using the discrete element method (DEM) \cite{doi:10.1680/geot.1979.29.1.47}.
In our DEM simulations, aggregates consist of a large number of monodisperse spherical ice particles with a radius of $r_{\rm p} = 100~\si{nm}$.
Aggregates consisting of submicron-sized ice particles are regarded as the building blocks of planets, and their collisional behavior has been investigated in several studies in the context of astrophysics \cite{2021ApJ...915...22H, 2022MNRAS.509.5641S, 2023ApJ...951L..16A}.
Each particle is modeled as an elastic cohesive sphere, and the material parameters are equal to those assumed in previous studies \cite{2011ApJ...737...36W, 2023ApJ...951L..16A}.

We calculate the translational and rotational motions of each particle by solving the Newton--Euler equations.
The time derivatives of velocity and angular velocity of the $i$th particle, $\dot{\bm{v}_{i}}$ and $\dot{\bm{\omega}_{i}}$, are given by
\begin{equation}
\dot{\bm{v}_{i}} = \frac{1}{m_{\rm p}} \sum_{j} \bm{F}_{ij}
\end{equation}
and
\begin{equation}
\dot{\bm{\omega}_{i}} = \frac{1}{I_{\rm p}} \sum_{j} \bm{M}_{ij},
\end{equation}
where $m_{\rm p} = 4 \pi \rho_{\rm p} {r_{\rm p}}^{3} / 3$ and $I_{\rm p} = 2 m_{\rm p} {r_{\rm p}}^{2} / 5$ are the mass and moment of inertia of each particle, respectively.
The material density of ice is set to $\rho_{\rm p} = 1000~\si{kg.m^{-3}}$ \cite{1997ApJ...480..647D, 2007ApJ...661..320W}.
The force and torque on the $i$th particle exerted by the $j$th particle are denoted by $\bm{F}_{ij}$ and $\bm{M}_{ij}$, respectively.
The summations are taken for all particles in contact with the $i$th particle.

\subsection{Normal Force}

The interparticle normal force is described by a contact model for elastic cohesive spheres called Johnson--Kendall--Roberts (JKR) model \cite{1971RSPSA.324..301J}.
The JKR model has been widely used for DEM simulations of low-velocity collisions of ice aggregates \cite{1997ApJ...480..647D, 2013A&A...559A..62W, 2021ApJ...915...22H}.\footnote{
We acknowledge, however, that molecular dynamics simulations have reported that plastic deformation and melting of ice occur when the interparticle relative velocity exceeds several hundred m/s \cite{2022NatSR..1213858N}.}
The normal force on the particle $1$ exerted by the particle $2$, $\bm{F}_{{\rm n}, 12}$, is
\begin{equation}
\bm{F}_{{\rm n}, 12} = F_{\rm p} \bm{n}_{\rm c},
\end{equation}
where $\bm{n}_{\rm c}$ is a unit vector defined by
\begin{equation}
\bm{n}_{\rm c} = \frac{\bm{x}_{1} - \bm{x}_{2}}{|| \bm{x}_{1} - \bm{x}_{2} ||}.
\end{equation}
The position of the particle $1$ is $\bm{x}_{1}$, $|| \bm{u} ||$ denotes the norm of the vector $\bm{u}$, and $F_{\rm p}$ is the normal force acting between particles 1 and 2.
We note that $F_{\rm p}$ is positive when the repulsive force works.

At an equilibrium state where $F_{\rm p} = 0$, the interparticle contact radius, $a_{\rm p}$, is equal to the equilibrium radius, $a_{0} = {[ {( 9 \pi \gamma {r_{\rm p}}^{2} )} / {( 4 E_{\rm p}^{*} )} ]}^{1/3}$, where $\gamma = 0.1~\si{J.m^{-2}}$ is the surface energy and $E_{\rm p}^{*} = E_{\rm p} / {[ 2 {( 1 - {\nu_{\rm p}}^{2} )} ]}$ is the reduced Young's modulus for interparticle contacts.
For ice spheres, the Young's modulus is $E_{\rm p} = 7~\si{GPa}$ and the Poisson's ratio is $\nu_{\rm p} = 0.25$ \cite{1997ApJ...480..647D, 2007ApJ...661..320W}.
The normal force acting between two particles is a function of the interparticle contact radius:
\begin{equation}
\frac{F_{\rm p}}{F_{\rm c}} = 4 {\left[ {\left( \frac{a_{\rm p}}{a_{0}} \right)}^{3} - {\left( \frac{a_{\rm p}}{a_{0}} \right)}^{3/2} \right]},
\end{equation}
where $F_{\rm c} = 3 \pi \gamma r_{\rm p} / 2$ is the maximum force needed to separate the two particles in contact.
The contact radius $a_{\rm p}$ is related to the interparticle compression length, $\delta_{\rm p}$, as follows:
\begin{equation}
\frac{\delta_{\rm p}}{\delta_{0}} = 3 {\left( \frac{a_{\rm p}}{a_{0}} \right)}^{2} - 2 {\left( \frac{a_{\rm p}}{a_{0}} \right)}^{1/2},
\end{equation}
where $\delta_{0} = {( 2 {a_{0}}^{2} )} / {( 3 r_{\rm p} )}$ is the compression length at the equilibrium state.
The compression length is given by
\begin{equation}
\delta_{\rm p} = 2 r_{\rm p} - {|| \bm{x}_{1} - \bm{x}_{2} ||}.
\end{equation}

\begin{figure}
\centering
\includegraphics[width = \columnwidth]{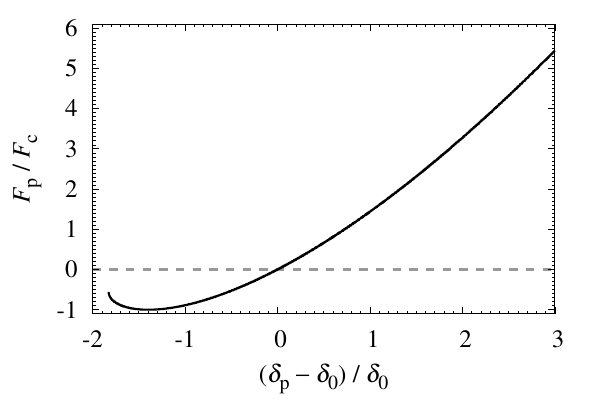}
\caption{
Normal force acting between two particles in contact, $F_{\rm p}$, with respect to the deviation from the equilibrium compression length, $\delta_{\rm p} - \delta_{0}$.
The maximum force needed to separate the two contact particles is $F_{\rm c}$, and $\delta_{0}$ is the compression length at the equilibrium state.
}
\label{fig_JKR}
\end{figure}

Therefore, the interparticle normal force is a function of the compression length.
Figure \ref{fig_JKR} shows $F_{\rm p}$ with respect to ${( \delta_{\rm p} - \delta_{0} )}$.
At the moment of disconnection, $\delta_{\rm p}$ is equal to $- {( 9 / 16)}^{1/3} \delta_{0}$, whereas $\delta_{\rm p} = 0$ when two particles connect.

\subsection{Tangential Motion}

The tangential motion of two particles in contact can be divided into three components: sliding, rolling, and twisting.
The displacements corresponding to tangential motions are expressed by the rotation of the two particles.
The forces ($\bm{F}$) and torques ($\bm{M}$) acting on the $i$th particle due to sliding (s), rolling (r), and twisting (t) with the $j$th particle are denoted by $\bm{F}_{{\rm s}, ij}$, $\bm{F}_{{\rm r}, ij}$, $\bm{F}_{{\rm t}, ij}$, $\bm{M}_{{\rm s}, ij}$, $\bm{M}_{{\rm r}, ij}$, and $\bm{M}_{{\rm t}, ij}$, respectively.
We note that $\bm{F}_{{\rm r}, ij} = \bm{F}_{{\rm t}, ij} = 0$ \cite{2007ApJ...661..320W}, and $\bm{F}_{ij}$ and $\bm{M}_{ij}$ are given by
\begin{equation}
\bm{F}_{ij} = \bm{F}_{{\rm n}, ij} + \bm{F}_{{\rm s}, ij}
\end{equation}
and
\begin{equation}
\bm{M}_{ij} = \bm{M}_{{\rm s}, ij} + \bm{M}_{{\rm r}, ij} + \bm{M}_{{\rm t}, ij},
\end{equation}
respectively.

The tangential interaction model used in this study is identical to that developed by Wada et al.~\cite{2007ApJ...661..320W}.
We assume that tangential motions are modeled as linear elastic springs when all displacements of the tangential motions are smaller than the thresholds.
We also consider the inelastic behavior when the displacements exceed the threshold values.
Details of the tangential interaction model are described in Appendix \ref{app:tangential}.

\section{Simulation Setup}
\label{sec:setup}

In this study, we use spherical aggregates with a radius of $R_{\rm agg} = 40$--$80 r_{\rm p}$, corresponding to $4$--$8~\si{\micro m}$.
The number of constituent particles in an aggregate, $N$, is given by $N = \phi_{\rm agg} {( R_{\rm agg} / r_{\rm p} )}^{3}$, where $\phi_{\rm agg}$ is the volume filling factor.
We focus on collisions of equal-size aggregates, and the total number of particles in a simulation is $2 N$.
In both numerical simulations and laboratory experiments, aggregates with $\phi_{\rm agg} \simeq 0.3$--$0.4$ have been widely used to investigate their bouncing behavior \cite{2012Icar..218..688W, 2023ApJ...951L..16A, TOYODA2024115964}.
In this study, we set $\phi_{\rm agg} = 0.3$ or $0.4$.

To create spherical dust aggregates, we employ the close-packing and particle-extraction (CPE) procedure \cite{2011ApJ...737...36W, 2023ApJ...951L..16A}.
The preparation procedure involves the following steps: an infinitely large cubic close-packed aggregate is initially created, and the center of a sphere with a radius of $R_{\rm agg}$ is randomly chosen.
Particles whose centers lie outside of the sphere are then extracted.
Additionally, particles are randomly removed from the aggregate until $N$ decreases to $\phi_{\rm agg} {( R_{\rm agg} / r_{\rm p} )}^{3}$.
Finally, the orientation of the aggregate is randomly selected.

\begin{figure}
\centering
\includegraphics[width = \columnwidth]{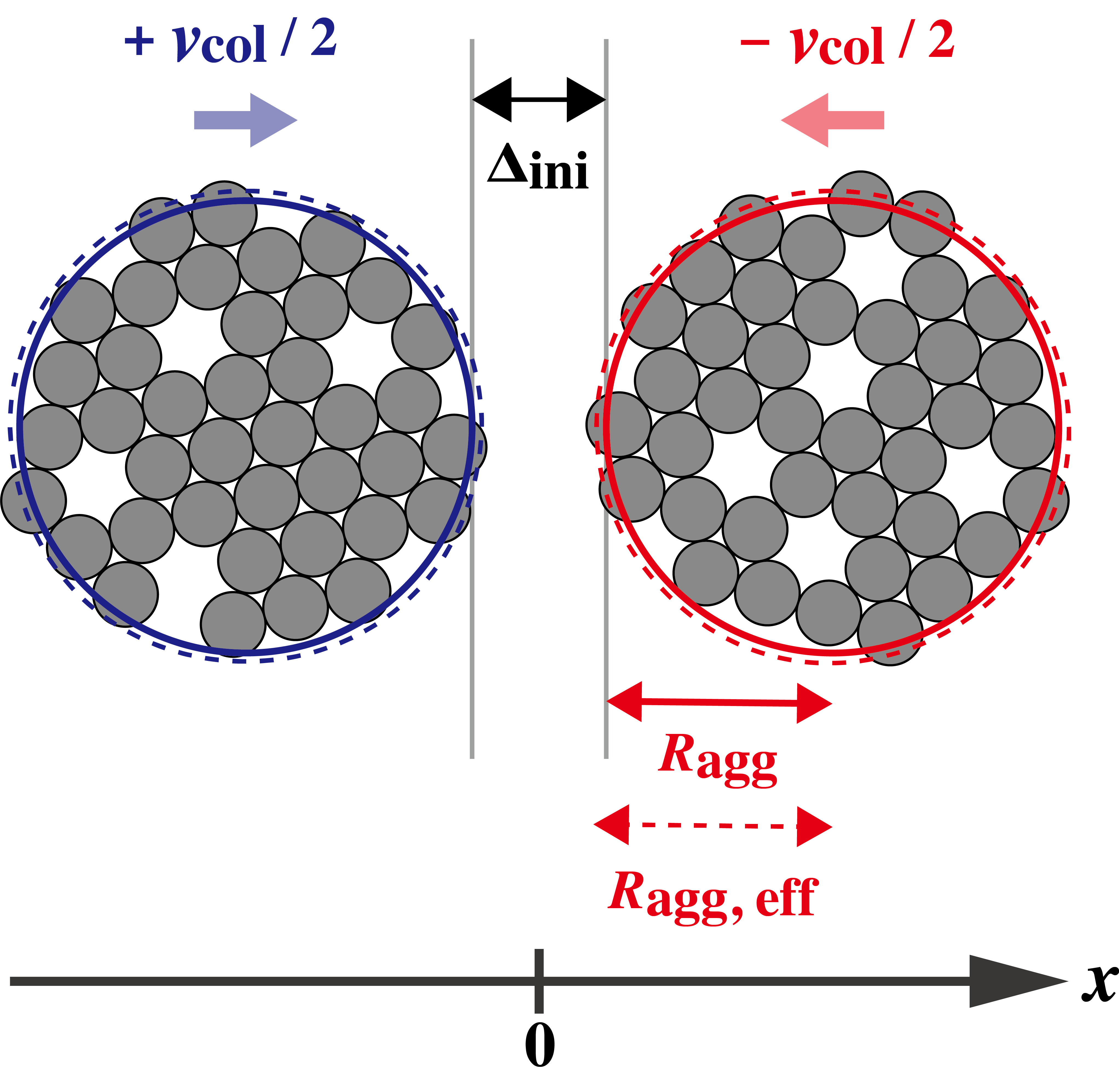}
\caption{
Schematic of the initial setup for numerical simulations.
Two equal-size aggregates with a radius of $R_{\rm agg}$ are prepared by CPE procedure, and we randomly choose the orientation of the aggregates.
The initial separation between two aggregates is $\Delta_{\rm ini} = 2 r_{\rm p}$, and the collision velocity is set to $v_{\rm col}$.
We also introduce the effective radius of aggregates, $R_{\rm agg, eff} = R_{\rm agg} + \delta_{\rm offset}$, where $\delta_{\rm offset}$ represents an offset value obtained from the linear dependence of the maximum compression length on the aggregate radius (see Section \ref{sec:size}).
}
\label{fig_setup}
\end{figure}

The schematic of the initial setup for numerical simulations is illustrated in Figure \ref{fig_setup}.
The initial separation between two aggregates is $\Delta_{\rm ini} = 2 r_{\rm p}$.
We position the centers of aggregates initially at ${( x, y, z )} = {( \pm {( R_{\rm agg} + r_{\rm p} )}, 0, 0 )}$.
The collision velocity of the two aggregates is set to $v_{\rm col} = 10^{0.1 i}~\si{m.s^{-1}}$, where $i = 0$, $1$, ..., $7$.
For each parameter set ${( \phi_{\rm agg}, R_{\rm agg}, v_{\rm col} )}$, we conduct 10 runs with different aggregates.

\section{Results}
\label{sec:results}

In this section, we show the numerical results and compare them with analytical model calculations.
We find that the interaggregate motion during collisional compression is well approximated by that of elastoplastic spheres.

\subsection{Temporal Evolution}

\begin{figure}
\centering
\includegraphics[width = \columnwidth]{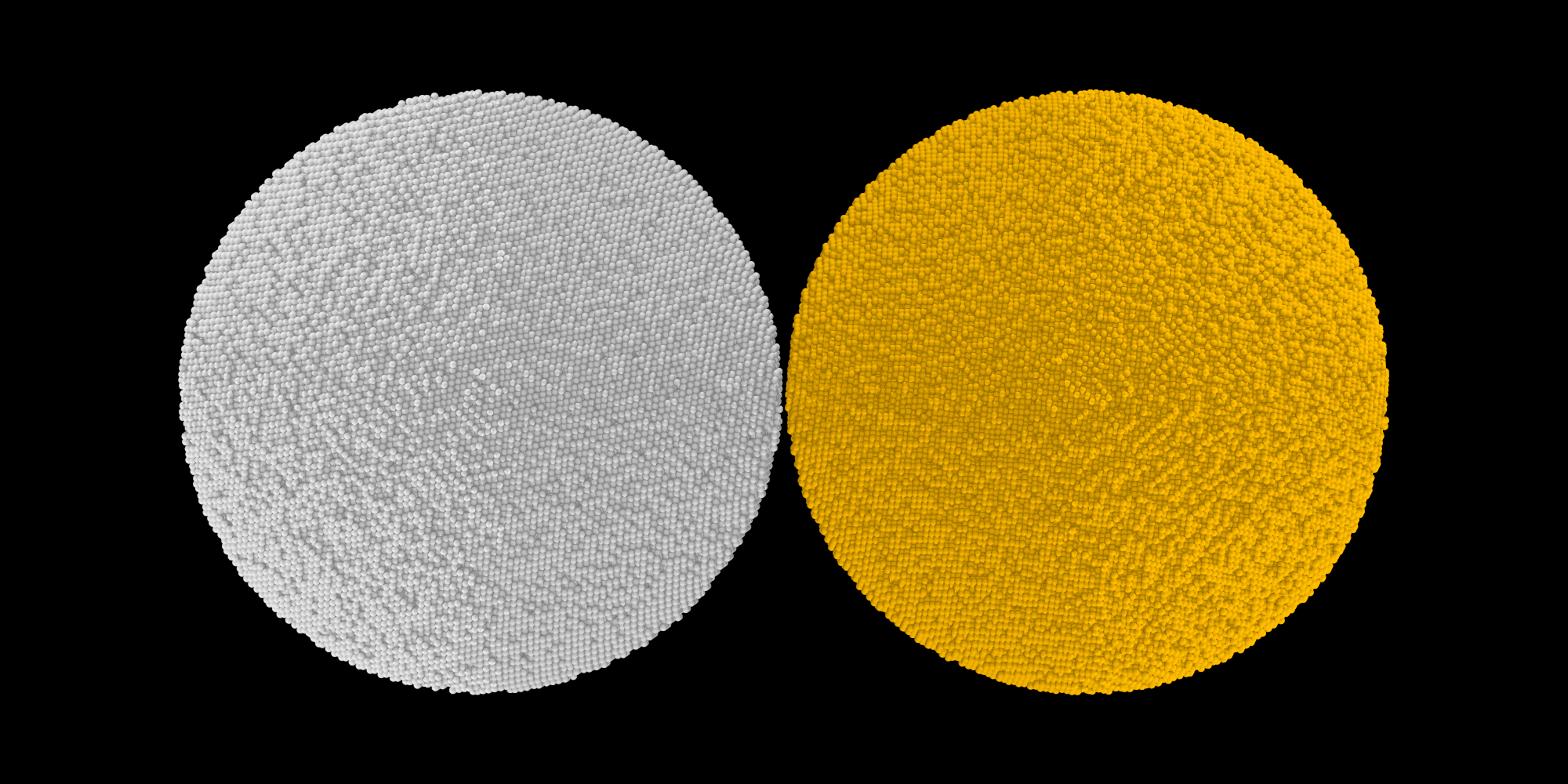}
\includegraphics[width = \columnwidth]{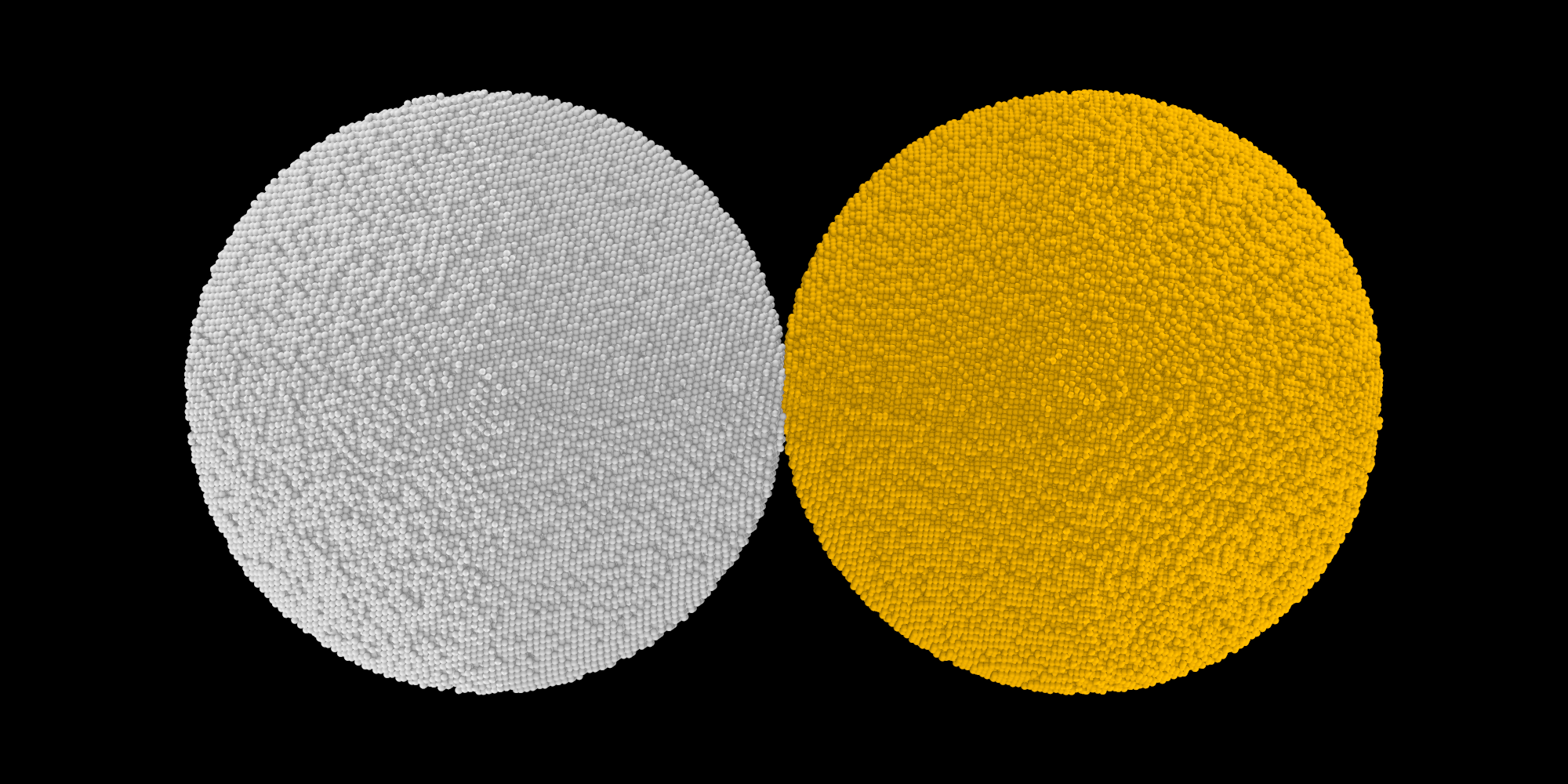}
\includegraphics[width = \columnwidth]{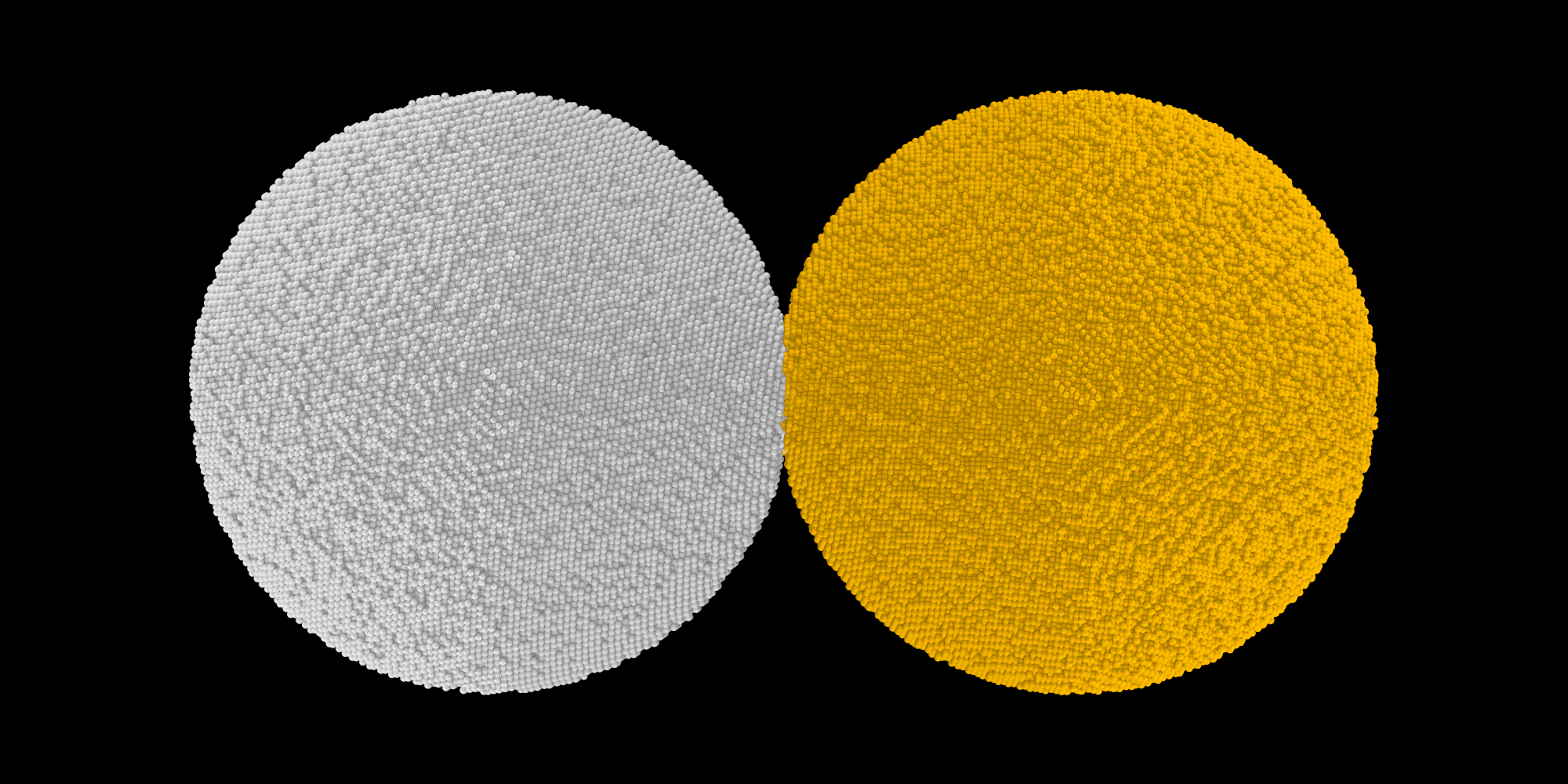}
\includegraphics[width = \columnwidth]{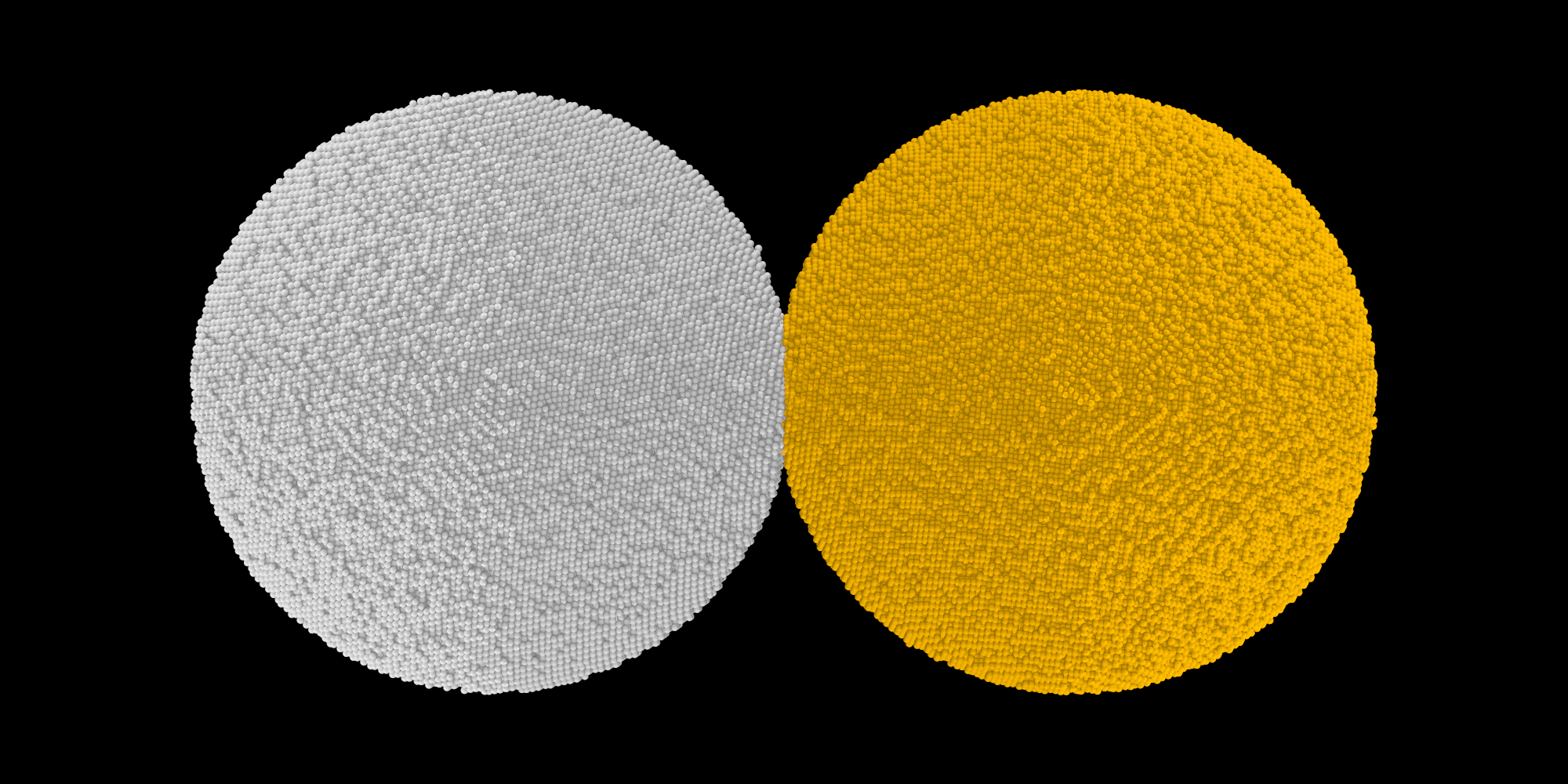}
\caption{
Snapshots of a collisional outcome for $\phi_{\rm agg} = 0.4$, $R_{\rm agg} = 80 r_{\rm p}$, and $v_{\rm col} = 3.2~\si{m.s^{-1}}$.
The first snapshot was taken at $t = 0$, and the time interval is $100~\si{ns}$.
}
\label{fig_snapshot}
\end{figure}

\begin{figure}
\centering
\includegraphics[width = \columnwidth]{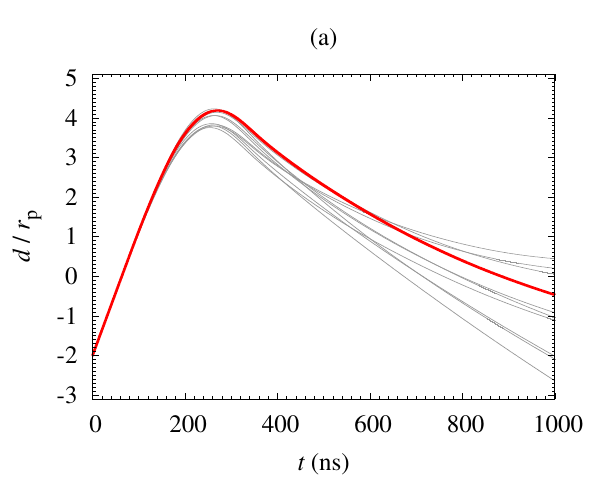}
\includegraphics[width = \columnwidth]{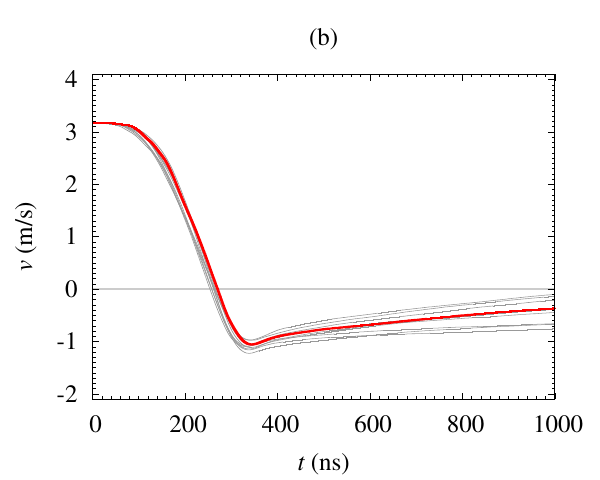}
\caption{
Temporal evolution of the compression length and the mutual velocity for $\phi_{\rm agg} = 0.4$, $R_{\rm agg} = 80 r_{\rm p}$, and $v_{\rm col} = 3.2~\si{m.s^{-1}}$.
The red lines show results for the run illustrated in Figure \ref{fig_snapshot}, and the grey lines are for the other 9 runs.
(a) Compression length.
(b) Mutual velocity.
}
\label{fig_temporal}
\end{figure}

We show numerical results for $\phi_{\rm agg} = 0.4$, $R_{\rm agg} = 80 r_{\rm p}$, and $v_{\rm col} = 3.2~\si{m.s^{-1}}$.
Figure \ref{fig_snapshot} shows snapshots of a collisional outcome.
The first snapshot was taken at $t = 0$, and the time interval is $100~\si{ns}$.
Two aggregates reach the maximum compression state at $t \simeq 300~\si{ns}$, and they still stick together at the end of the simulation ($t = 1000~\si{ns}$).
The red lines in Figures \ref{fig_temporal}(a) and \ref{fig_temporal}(b) show the temporal evolution of the compression length and the mutual velocity, $d$ and $v$, respectively.
In our DEM simulations, we define $d$ as
\begin{equation}
d \equiv 2 R_{\rm agg} - {\left( x_{+} - x_{-} \right)},
\end{equation}
where $x_{+}$ represents the x-component of the center of mass for particles with $x_{i} > 0$:
\begin{equation}
x_{+} \equiv \frac{1}{n_{+}} \sum_{x_{i} > 0} x_{i},
\end{equation}
with $n_{+}$ being the number of particles with $x_{i} > 0$.
Similarly, $v$ is defined as
\begin{equation}
v \equiv - {\left( v_{+} - v_{-} \right)},
\end{equation}
where $v_{+}$ denotes the x-component of the mean velocity of particles with $x_{i} > 0$:
\begin{equation}
v_{+} \equiv \frac{1}{n_{+}} \sum_{x_{i} > 0} v_{i}.
\end{equation}
We also define $x_{-}$ and $v_{-}$ in the same manner.
It is evident that $v = - {\rm d}d / {\rm d}{t}$, and $d$ reaches its maximum when $v = 0$.
The maximum value of $d$ is $d_{\rm max} \simeq 4 r_{\rm p}$.
We define $t_{\rm max}$ as the time when $d = d_{\rm max}$.

The collisional outcomes depend on the microstructure of the randomly prepared aggregates, even when macroscopic parameters (e.g., $\phi_{\rm agg}$, $R_{\rm agg}$, and $v_{\rm col}$) are fixed \cite{2011ApJ...737...36W, 2013A&A...551A..65S, 2023ApJ...951L..16A}.
To characterize the typical behavior of aggregates upon collision, we conduct 10 runs for each set of ${( \phi_{\rm agg}, R_{\rm agg}, v_{\rm col} )}$.
The gray lines in Figures \ref{fig_temporal}(a) and \ref{fig_temporal}(b) show the temporal evolution of $d$ and $v$ for the other 9 runs.
We observe variations in both $d$ and $v$ among the runs, particularly noticeable in the late separation stage ($t \gtrsim 400~\si{ns}$), where $v < 0$ and ${\rm d}v / {\rm d}t > 0$.
In other words, the aggregate--aggregate interaction in the late separation stage is stochastic.
In contrast, the variations of $d$ and $v$ around $t_{\rm max}$ are significantly smaller than those observed in the late stage.
Thus, we investigate the collisional compression process for $t \le t_{\rm max}$ using a (deterministic) mechanical model for macroscopic spheres (Section \ref{sec:theoretical}).

\subsection{Numerical Results}
\label{sec:numerical}

We investigate the maximum compression length, $\delta_{\rm max}$, and the time at maximum compression, $\tau_{\rm max}$.
In Section \ref{sec:numerical}, we present the results of our DEM simulations.
Theoretical interpretations of the numerical results are provided in Section \ref{sec:theoretical}.

\subsubsection{Size Dependence}
\label{sec:size}

\begin{figure}
\centering
\includegraphics[width = \columnwidth]{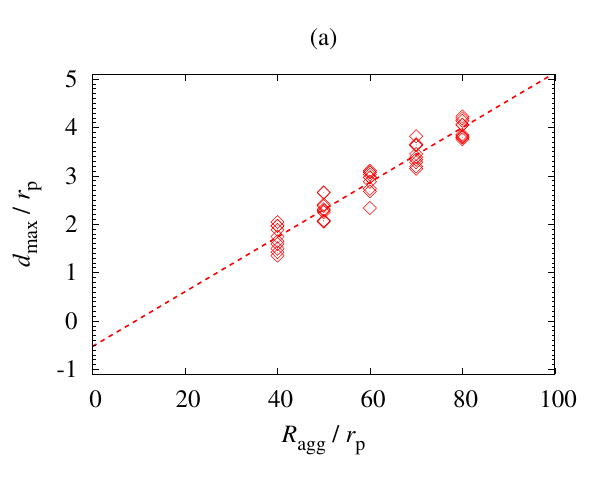}
\includegraphics[width = \columnwidth]{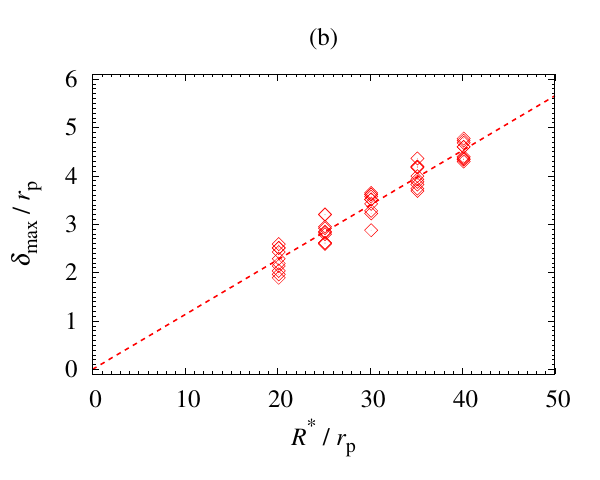}
\caption{
Size dependence of the maximum compression length for $\phi_{\rm agg} = 0.4$ and $v_{\rm col} = 3.2~\si{m.s^{-1}}$.
(a) $d_{\rm max}$ with respect to $R_{\rm agg}$.
(b) $\delta_{\rm max}$ with respect to $R^{*}$.
The scatter points are for different simulation runs.
Dashed lines represent the best fits obtained by the least squares method.
}
\label{fig_delta_Rstar}
\end{figure}

The size dependence of the maximum compression length is crucial for understanding the mechanical properties of aggregates as macroscopic spheres.
Figure \ref{fig_delta_Rstar}(a) illustrates the relationship between $d_{\rm max}$ and $R_{\rm agg}$ for $\phi_{\rm agg} = 0.4$ and $v_{\rm col} = 3.2~\si{m.s^{-1}}$.
Notably, $d_{\rm max}$ appears to follow a linear function of $R_{\rm agg}$.
The dashed line in Figure \ref{fig_delta_Rstar}(a) represents the best fit obtained through the least squares method:
\begin{equation}
d_{\rm max} = c_{1} R_{\rm agg} - c_{2} r_{\rm p},
\end{equation}
where the fitting coefficients are $c_{1} = 5.66 \times 10^{-2}$ and $c_{2} = 0.527$.

The maximum compression length becomes proportional to the size of aggregates when we introduce an offset value.
In this context, we introduce the effective radius of aggregates, $R_{\rm agg, eff}$ (see Figure \ref{fig_setup}), defined as
\begin{equation}
R_{\rm agg, eff} \equiv R_{\rm agg} + \delta_{\rm offset},
\end{equation}
where $\delta_{\rm offset}$ represents the offset of the effective surface relative to that in the CPE procedure.
The reduced aggregate radius, $R^{*}$, is given by  
\begin{equation}
R^{*} = \frac{R_{\rm agg, eff}}{2},
\end{equation}
and the effective compression length, $\delta$, is defined as
\begin{equation}
\delta \equiv d + 2 \delta_{\rm offset}.
\end{equation}
Then, the maximum compression length is given by $\delta_{\rm max} = d_{\rm max} + 2 \delta_{\rm offset}$.

Figure \ref{fig_delta_Rstar}(b) shows the dependence of $\delta_{\rm max}$ on $R^{*}$ for $\phi_{\rm agg} = 0.4$ and $v_{\rm col} = 3.2~\si{m.s^{-1}}$.
It becomes evident that $\delta_{\rm max}$ is proportional to $R^{*}$ when an appropriate $\delta_{\rm offset}$ value is chosen, given by
\begin{eqnarray}
\delta_{\rm offset} & = & \frac{c_{2} r_{\rm p}}{2 - c_{1} / 2} \nonumber \\
                    & = & 0.271 r_{\rm p}.
\end{eqnarray}
The dashed line in Figure \ref{fig_delta_Rstar}(b) represents the best fit:
\begin{eqnarray}
\delta_{\rm max} & = & 2 c_{1} R^{*} \nonumber \\
                 & = & 0.113 R^{*}.
\label{eq:delta_Rstar}
\end{eqnarray}

The value of $\delta_{\rm offset}$ obtained from our DEM simulations falls within a reasonable range.
When preparing CPE aggregates, we extract particles whose center is outside of the sphere with a radius of $R_{\rm agg}$ (see Section \ref{sec:setup}).
Therefore, the sphere with a radius of $R_{\rm agg} + r_{\rm p}$ includes all constituent particles.
While we can imagine that $R_{\rm agg, eff}$ would be close to $R_{\rm agg}$, it must be smaller than $R_{\rm agg} + r_{\rm p}$.
The chosen value of $\delta_{\rm offset} = 0.271 r_{\rm p}$ satisfies these conditions.

\begin{figure}
\centering
\includegraphics[width = \columnwidth]{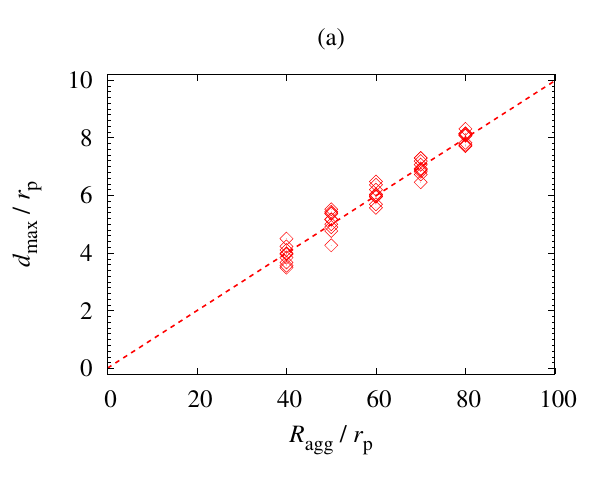}
\includegraphics[width = \columnwidth]{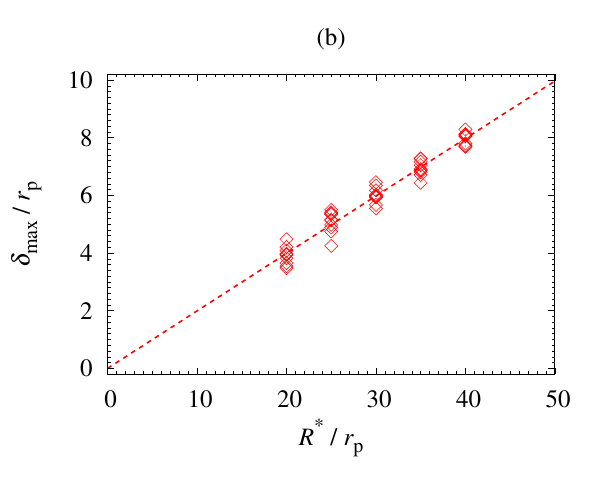}
\caption{
Same as Figure \ref{fig_delta_Rstar}, but for $\phi_{\rm agg} = 0.3$ and $v_{\rm col} = 3.2~\si{m.s^{-1}}$.
}
\label{fig_delta_Rstar_phi30}
\end{figure}

We note that the $\delta_{\rm offset}$ and $\delta_{\rm max}$ values reported above are only applicable to aggregates with $\phi_{\rm agg} = 0.4$ and depends on $\phi_{\rm agg}$.
Figure \ref{fig_delta_Rstar_phi30} shows the size dependence of the maximum compression length for $\phi_{\rm agg} = 0.3$ and $v_{\rm col} = 3.2~\si{m.s^{-1}}$.
We find that
\begin{equation}
\delta_{\rm offset} = - 0.010 r_{\rm p}
\end{equation}
and
\begin{equation}
\delta_{\rm max} = 0.199 R^{*}.
\label{eq:delta_Rstar_phi30}
\end{equation}
It is evident that $\delta_{\rm max}$ for $\phi_{\rm agg} = 0.3$ is larger than that for $\phi_{\rm agg} = 0.4$ when $R^{*}$ is fixed.

\begin{figure}
\centering
\includegraphics[width = \columnwidth]{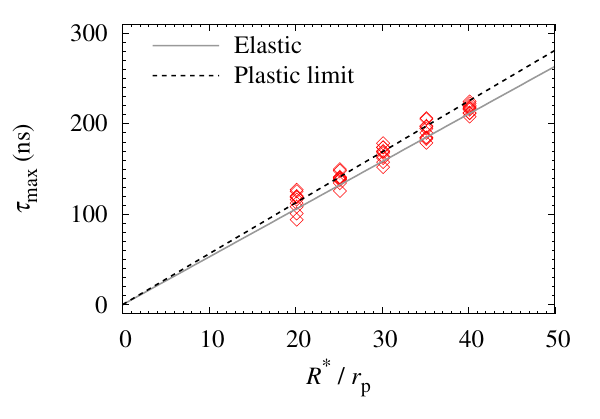}
\caption{
Size dependence of the time at maximum compression, $\tau_{\rm max}$, for $\phi_{\rm agg} = 0.4$ and $v_{\rm col} = 3.2~\si{m.s^{-1}}$.
The gray solid line represents the analytical result for the perfectly elastic case, whereas the black dashed line represents the rigid plastic limit.
}
\label{fig_tau_Rstar}
\end{figure}

We also investigate the time at maximum compression.
Inter-aggregate contacts initiate when $\delta = 0$, and the time delay from the start of numerical integration is given by $t_{\rm col} = {( \Delta_{\rm ini} - 2 \delta_{\rm offset} )} / v_{\rm col}$.
We define the time from the onset of effective contact, $\tau$:
\begin{equation}
\tau \equiv t - t_{\rm col},
\end{equation}
and we define $\tau_{\rm max} \equiv \tau|_{\delta = \delta_{\rm max}}$ as the time at the maximum compression.
If $d$ reaches its maximum $d_{\rm max}$ at $t = t_{\rm max}$, then $\tau_{\rm max}$ is given by $\tau_{\rm max} = t_{\rm max} - t_{\rm col}$.
The red diamonds in Figure \ref{fig_tau_Rstar} illustrates $\tau_{\rm max}$ with respect to $R^{*}$, for $\phi_{\rm agg} = 0.4$ and $v_{\rm col} = 3.2~\si{m.s^{-1}}$.
Notably, $\tau_{\rm max}$ demonstrates a proportional relationship with $R^{*}$.
This linear relationship is consistent with the theoretical prediction for elastoplastic spheres \cite{doi:10.1080/14786443008565033} (see Section \ref{sec:theoretical}).

\subsubsection{Velocity Dependence}
\label{sec:velocity}

In Section \ref{sec:size}, we showed results for a fixed collision velocity of $v_{\rm col} = 3.2~\si{m.s^{-1}}$ while investigating the size dependence.
In this section, we present the velocity dependence of $\delta_{\rm max}$ and $\tau_{\rm max}$.

\begin{figure*}
\centering
\includegraphics[width = \columnwidth]{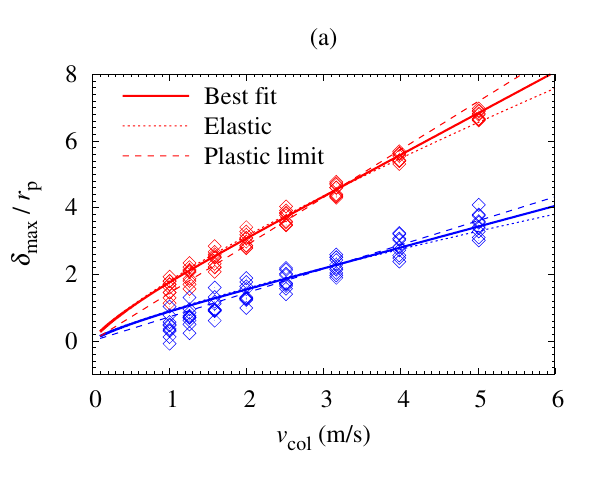}
\includegraphics[width = \columnwidth]{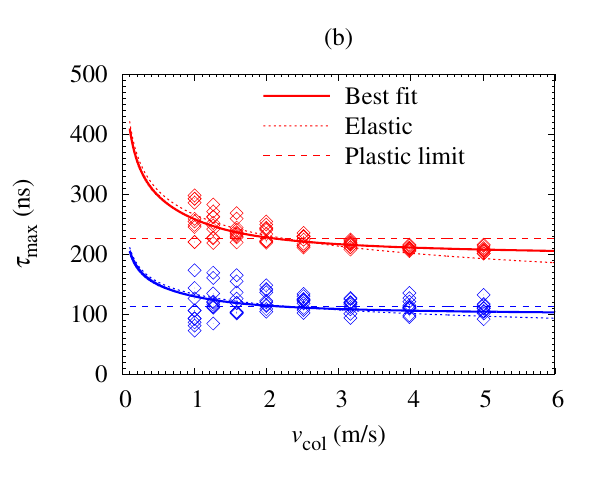}
\caption{
Velocity dependence of $\delta_{\rm max}$ and $\tau_{\rm max}$ for $\phi_{\rm agg} = 0.4$.
(a) $\delta_{\rm max}$ with respect to $v_{\rm col}$.
(b) $\tau_{\rm max}$ with respect to $v_{\rm col}$.
The red and blue diamonds denote the numerical results for $R_{\rm agg} = 80 r_{\rm p}$ and $40 r_{\rm p}$, respectively.
The scatter points are for different simulation runs.
Dotted lines show the analytical results for the perfectly elastic case, whereas dashed lines are those for the rigid plastic limit.
Solid lines represent the analytical results for the best-fit model with $\sigma_{\rm plastic} = 2.25~\si{MPa}$ and $E^{*} = 15.7~\si{MPa}$.
}
\label{fig_delta_tau_Vin}
\end{figure*}

Figure \ref{fig_delta_tau_Vin}(a) illustrates the variation of $\delta_{\rm max}$ with respect to $v_{\rm col}$ for $\phi_{\rm agg} = 0.4$.
The red and blue diamonds denote the numerical results for $R_{\rm agg} = 80 r_{\rm p}$ and $40 r_{\rm p}$, respectively.
We find that $\delta_{\rm max}$ increases with $v_{\rm col}$.
The dispersion for $R_{\rm agg} = 80 r_{\rm p}$ is smaller than that for $R_{\rm agg} = 40 r_{\rm p}$.
Figure \ref{fig_delta_tau_Vin}(b) shows the relationship between $\tau_{\rm max}$ and $v_{\rm col}$ for $\phi_{\rm agg} = 0.4$.
Despite a considerable dispersion, a discernible decreasing trend in $\tau_{\rm max}$ is observed, particularly for $R_{\rm agg} = 80 r_{\rm p}$.
Similar to the case in Figure \ref{fig_delta_tau_Vin}(a), the dispersion for $R_{\rm agg} = 80 r_{\rm p}$ appears to be smaller than that for $R_{\rm agg} = 40 r_{\rm p}$.

\begin{figure*}
\centering
\includegraphics[width = \columnwidth]{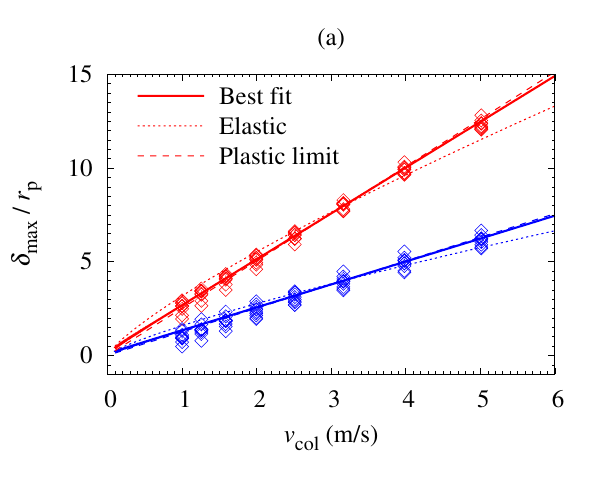}
\includegraphics[width = \columnwidth]{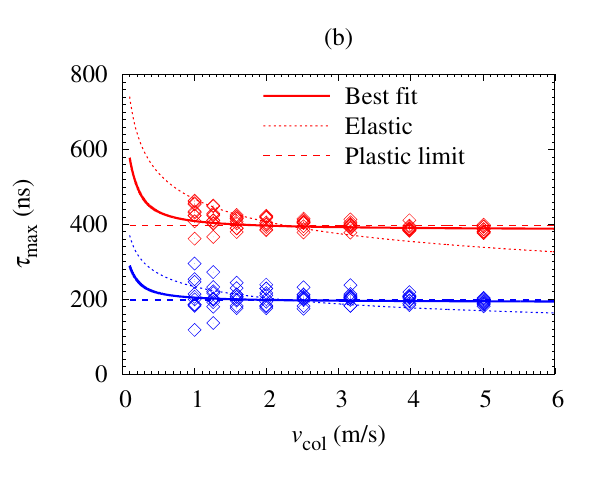}
\caption{
Same as Figure \ref{fig_delta_tau_Vin}, but for $\phi_{\rm agg} = 0.3$.
Solid lines represent the analytical results for the best-fit model with $\sigma_{\rm plastic} = 0.428~\si{MPa}$ and $E^{*} = 4.96~\si{MPa}$.
}
\label{fig_delta_tau_Vin_phi30}
\end{figure*}

We also investigate the velocity dependence of $\delta_{\rm max}$ and $\tau_{\rm max}$ for $\phi_{\rm agg} = 0.3$.
We find that $\delta_{\rm max}$ increases with $v_{\rm col}$ (Figure \ref{fig_delta_tau_Vin_phi30}(a)), as in the case of $\phi_{\rm agg} = 0.4$.
In contrast to the case for $\phi_{\rm agg} = 0.4$, no clear decreasing trend in $\tau_{\rm max}$ is observed for $\phi_{\rm agg} = 0.3$ (Figure \ref{fig_delta_tau_Vin_phi30}(b)).

We see a large variation in $\tau_{\rm max}$ among simulation runs for low-velocity collisions (Figures \ref{fig_delta_tau_Vin}(b) and \ref{fig_delta_tau_Vin_phi30}(b)).
The variation in $\delta_{\rm max}$, normalized by its average, is also large for low-velocity collisions (Figures \ref{fig_delta_tau_Vin}(a) and \ref{fig_delta_tau_Vin_phi30}(a)).
The compressed area during collision is small for low-velocity collisions, resulting in large variations reflecting the local heterogeneity.

\subsection{Theoretical Interpretation}
\label{sec:theoretical}

In Section \ref{sec:numerical}, we found that $\delta_{\rm max}$ is proportional to $R^{*}$.
We can explain the linear dependence of $\delta_{\rm max}$ on $R^{*}$ when dust aggregates behave as elastoplastic spheres.
In this section, we derive the size and velocity dependence of the maximum compression length using a theoretical model \cite{doi:10.1080/14786443008565033}.
Additionally, we provide constraints on the mechanical properties of aggregates.
Further details of the contact model are provided in Appendix \ref{app:Andrews}.

Andrews \cite{doi:10.1080/14786443008565033} considered a simple contact model where, if $\delta$ is smaller than the critical compression length for yielding, $\delta_{\rm crit}$, two spheres are assumed to behave as perfectly elastic spheres.
In contrast, elastoplastic deformation occurs when $\delta > \delta_{\rm crit}$.

For $\delta \le \delta_{\rm crit}$, the repulsive force, $F = F {( \delta )}$, follows the Hertzian model \cite{hertz1896miscellaneous} (Equation (\ref{eq:F_elastic_app})):
\begin{equation}
F {( \delta )} = \frac{4 E^{*} \sqrt{R^{*}} \delta^{3/2}}{3},
\end{equation}
where $E^{*}$ is the reduced Young's modulus.
The critical compression length for yielding is given by (Equation (\ref{eq:delta_crit}))
\begin{equation}
\delta_{\rm crit} = {\left( \frac{\pi \sigma_{\rm plastic}}{2 E^{*}} \right)}^{2} R^{*},
\label{eq:delta_crit_main}
\end{equation}
where $\sigma_{\rm plastic}$ is the yield stress, representing the critical normal stress for plastic deformation.
For $\delta_{\rm max} \le \delta_{\rm crit}$, the total work done by the repulsive force during compression is
\begin{eqnarray}
W_{\rm comp} & = & - \int_{0}^{\delta_{\rm max}}~F~{\rm d}\delta \nonumber \\
             & = & - \frac{8 E^{*} \sqrt{R^{*}} {\delta_{\rm max}}^{5/2}}{15}.
\label{eq:W_elastic}
\end{eqnarray}

In contrast, for $\delta > \delta_{\rm crit}$, elastoplastic deformation occurs.
Andrews \cite{doi:10.1080/14786443008565033} proposed that $F$ is given as follows (Equation (\ref{eq:F_plastic_app})):
\begin{equation}
F {( \delta )} = \pi R^{*} \sigma_{\rm plastic} {\left( \delta - \frac{1}{3} \delta_{\rm crit} \right)}.
\end{equation}
For $\delta_{\rm max} > \delta_{\rm crit}$, $W_{\rm comp}$ is given by
\begin{eqnarray}
W_{\rm comp} & = & - {\left( \int_{0}^{\delta_{\rm crit}}~F~{\rm d}\delta + \int_{\delta_{\rm crit}}^{\delta_{\rm max}}~F~{\rm d}\delta \right)} \nonumber \\
             & = & - \pi R^{*} \sigma_{\rm plastic} \nonumber \\
             &   & \cdot {\left( \frac{{\delta_{\rm crit}}^2}{10} - \frac{\delta_{\rm crit} \delta_{\rm max}}{3} + \frac{{\delta_{\rm max}}^{2}}{2} \right)}.
\label{eq:W_plastic}
\end{eqnarray}

We can calculate $\delta_{\rm max}$ by solving the following energy equation:
\begin{equation}
K_{\rm ini} + W_{\rm comp} = 0,
\end{equation}
where $K_{\rm ini}$ is the initial kinetic energy of relative motion.
Assuming that two aggregates have an equal radius of $R_{\rm agg, eff}$, $K_{\rm ini}$ is given by the following equation:
\begin{equation}
K_{\rm ini} = \frac{8 \pi}{3} \rho_{\rm agg} {v_{\rm col}}^{2} {R^{*}}^{3},
\label{eq:K}
\end{equation}
where $\rho_{\rm agg} = \phi_{\rm agg} \rho_{\rm p}$ is the density of aggregates.

\subsubsection{Size Dependence}
\label{sec:size_theoretical}

It can be shown that $\delta_{\rm max}$ given by Equations (\ref{eq:W_plastic})--(\ref{eq:K}) is in general proportional to $R^{*}$.
For $\delta_{\rm max} \le \delta_{\rm crit}$, $\delta_{\rm max}$ is derived from Equations (\ref{eq:W_elastic}) and (\ref{eq:K}) and is given by \cite{landau1986theory}
\begin{equation}
\delta_{\rm max} = {\left( 5 \pi \frac{\rho_{\rm agg}}{E^{*}} \right)}^{2/5} {v_{\rm col}}^{4/5} R^{*},
\label{eq:delta_elastic}
\end{equation}
whereas for the rigid plastic limit ($E^{*} \to \infty$ and $\delta_{\rm crit} \to 0$), $\delta_{\rm max}$ is derived from Equations (\ref{eq:W_plastic}) and (\ref{eq:K}) and is given by
\begin{equation}
\delta_{\rm max} = \frac{4}{\sqrt{3}} \sqrt{\frac{\rho_{\rm agg}}{\sigma_{\rm plastic}}} v_{\rm col} R^{*}.
\label{eq:delta_plastic}
\end{equation}

The interaggregate contact radius at the maximum compression, $a_{\rm max}$, is given by $a_{\rm max} = \sqrt{R^{*} \delta_{\rm max}}$, which is also proportional to $R^{*}$.
Therefore, the volume of the compressed region is proportional to ${R^{*}}^{3}$.

\subsubsection{Relation between $\sigma_{\rm plastic}$ and $E^{*}$}
\label{sec_sigma_and_E_theoretical}

The normalized maximum compression length $\delta_{\rm max} / R^{*}$ determined by Equations (\ref{eq:W_plastic})--(\ref{eq:K}) depends on the macroscopic parameters $\sigma_{\rm plastic}$ and $E^{*}$.
However, the values of these parameters are unconstrained so far.
Below, we constrain $\sigma_{\rm plastic}$ and $E^{*}$ by using our simulation results: $\delta_{\rm max} / R^{*} = 0.113$ for $\phi_{\rm agg} = 0.4$ and $v_{\rm col} = 3.2~\si{m.s^{-1}}$ (Equation (\ref{eq:delta_Rstar})), and $\delta_{\rm max} / R^{*} = 0.199$ for $\phi_{\rm agg} = 0.3$ and $v_{\rm col} = 3.2~\si{m.s^{-1}}$ (Equation (\ref{eq:delta_Rstar_phi30})).

\begin{figure*}
\centering
\includegraphics[width = \columnwidth]{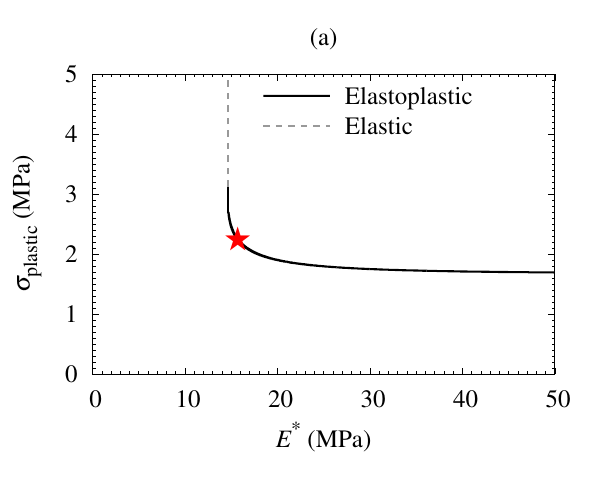}
\includegraphics[width = \columnwidth]{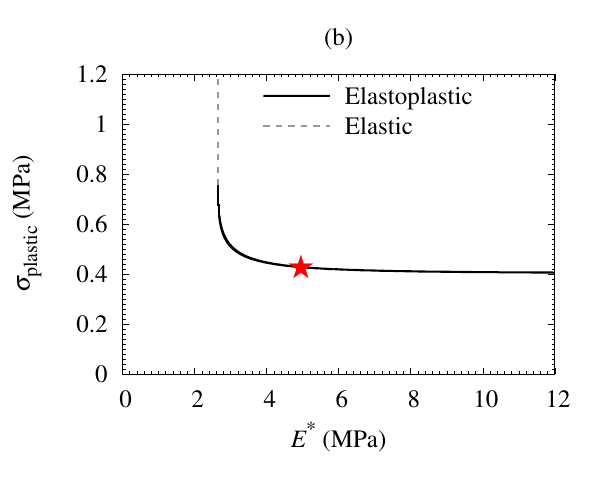}
\caption{
Pairs of ${( \sigma_{\rm plastic}, E^{*} )}$ that can explain the $\delta_{\rm max} / R^{*}$ value at $v_{\rm col} = 3.2~\si{m.s^{-1}}$ (Equation (\ref{eq:delta_Rstar})).
(a) For $\phi_{\rm agg} = 0.4$.
(b) For $\phi_{\rm agg} = 0.3$.
The black solid line represents the pair of ${( \sigma_{\rm plastic}, E^{*} )}$ for the elastoplastic case, whereas the gray dashed line is that for the elastic case.
The red star represents the best-fit parameters obtained from the velocity dependence (see Section \ref{sec:velocity_theoretical}).
}
\label{fig_Estar_sigma}
\end{figure*}

Figure \ref{fig_Estar_sigma}(a) shows pairs of ${( \sigma_{\rm plastic}, E^{*} )}$ that reproduces this simulation result for $\phi_{\rm agg}  =0.4$.
It can be shown from Equations (\ref{eq:delta_crit_main}) and (\ref{eq:delta_elastic}) that aggregates behave elastically during compression, i.e., $\delta_{\rm crit} > \delta_{\rm max}$, when $\sigma_{\rm plastic} > 3.13~\si{MPa}$.
In this case, the corresponding $E^{*}$ that can explain the dependence shown in Figure \ref{fig_delta_Rstar}(b) is derived from Equation (\ref{eq:delta_elastic}) as $E^{*} = 14.6~\si{MPa}$.
In contrast, $\delta_{\rm max} \ge \delta_{\rm crit}$ when $\sigma_{\rm plastic} \le 3.13~\si{MPa}$.
In this case, the pair of ${( \sigma_{\rm plastic}, E^{*} )}$ is determined by solving Equations (\ref{eq:delta_crit_main}), (\ref{eq:W_plastic}), and (\ref{eq:K}).
In the rigid plastic limit ($E^{*} \to \infty$ and $\delta_{\rm crit} \to 0$), $\sigma_{\rm plastic}$ attains the minimum value of $\sigma_{\rm plastic} = 1.67~\si{MPa}$.
We also shows pairs of ${( \sigma_{\rm plastic}, E^{*} )}$ that reproduces the simulation result for $\phi_{\rm agg} = 0.3$ in Figure \ref{fig_Estar_sigma}(b).




As shown in Figure \ref{fig_Estar_sigma}, a degeneracy between $\sigma_{\rm plastic}$ and $E^{*}$ exists.
This arises from the utilization of only one constraint, $\delta_{\rm max} / R^{*}$, with two parameters ($\sigma_{\rm plastic}$ and $E^{*}$).
In Section \ref{sec:velocity_theoretical}, we investigate the velocity dependence of $\delta_{\rm max}$ to resolve this degeneracy.
The star plotted in Figure \ref{fig_Estar_sigma} corresponds to the best-fit parameters derived from the velocity dependence.

Additionally, we try to give another constraint on the pair ${( \sigma_{\rm plastic}, E^{*} )}$ from the size dependence of $\tau_{\rm max}$.
It can be shown that $\tau_{\rm max}$ is in general proportional to $R^{*}$. 
When $\delta_{\rm max} \le \delta_{\rm crit}$, $\tau_{\rm max}$ is given by \cite{landau1986theory}
\begin{eqnarray}
\tau_{\rm max} & = & C_{\rm H} \frac{\delta_{\rm max}}{v_{\rm col}} \nonumber \\
               & = & 4.43 {\left( \frac{\rho_{\rm agg}}{E^{*}} \right)}^{2/5} {v_{\rm col}}^{- 1/5} R^{*},
\label{eq:tau_elastic}
\end{eqnarray}
where $C_{\rm H} = \int_{0}^{1} {( 1 - x^{5/2} )}^{- 1/2}~{\rm d}x = 1.47$.
For the rigid plastic limit ($E^{*} \to \infty$), in contrast, $\tau_{\rm max}$ is given by
\begin{eqnarray}
\tau_{\rm max} & = & \frac{\pi \delta_{\rm max}}{2 v_{\rm col}} \nonumber \\
               & = & \frac{2 \pi}{\sqrt{3}} \sqrt{\frac{\rho_{\rm agg}}{\sigma_{\rm plastic}}} R^{*}.
\label{eq:tau_plastic}
\end{eqnarray}
The gray solid line in Figure \ref{fig_tau_Rstar} represents the analytical result for the perfectly elastic case (Equation (\ref{eq:tau_elastic})), and the black dashed line is that for the rigid plastic limit (Equation (\ref{eq:tau_plastic})).
For $\phi_{\rm agg} = 0.4$, both elastic and plastic cases roughly match the numerical results, making it challenging to constrain the appropriate range of the pair ${( \sigma_{\rm plastic}, E^{*} )}$ from this result.

\subsubsection{Velocity Dependence}
\label{sec:velocity_theoretical}

In this section, we present the velocity dependence of $\delta_{\rm max}$ and $\tau_{\rm max}$, aiming to further constrain the pair ${( \sigma_{\rm plastic}, E^{*} )}$.
The dotted lines in Figures \ref{fig_delta_tau_Vin}(a) and \ref{fig_delta_tau_Vin}(b) represent the analytical results for the perfectly elastic case (i.e., $\sigma_{\rm plastic} = \infty$ and $E^{*} = 14.6~\si{MPa}$).
Correspondingly, the dashed lines depict results for the rigid plastic limit (i.e., $\sigma_{\rm plastic} = 1.67~\si{MPa}$ and $E^{*} = \infty$).
In the perfectly elastic case, $\delta_{\rm max}$ follows a proportional relationship with ${v_{\rm col}}^{4/5}$ (Equation (\ref{eq:delta_elastic})), and $\tau_{\rm max}$ is proportional to ${v_{\rm col}}^{- 1/5}$ (Equation (\ref{eq:tau_elastic})).
In contrast, for the rigid plastic limit, $\delta_{\rm max}$ is proportional to $v_{\rm col}$ (Equation (\ref{eq:delta_plastic})), whereas $\tau_{\rm max}$ remains independent of $v_{\rm col}$ (Equation (\ref{eq:tau_plastic})).
For $\phi_{\rm agg} = 0.4$, our numerical results (red and gray diamonds) fall within the range delineated by these two limiting cases.

We determine the optimal parameter set ${( \sigma_{\rm plastic}, E^{*} )}$ using the weighted least squares method.
Here, we use the velocity dependence of both $\delta_{\rm max}$ and $\tau_{\rm max}$, which are plotted on Figures \ref{fig_delta_tau_Vin}(a) and \ref{fig_delta_tau_Vin}(b).
For $\phi_{\rm agg} = 0.4$, the optimal parameter set obtained from our numerical results is
\begin{equation}
{\left( \sigma_{\rm plastic}, E^{*} \right)} = {\left( 2.25~\si{MPa}, 15.7~\si{MPa} \right)}.
\end{equation}
The solid lines in Figures \ref{fig_delta_tau_Vin}(a) and \ref{fig_delta_tau_Vin}(b) represent the analytical results for this best-fit model.
The star plotted in Figure \ref{fig_Estar_sigma}(a) corresponds to the pair ${( \sigma_{\rm plastic}, E^{*} )}$ for the best fit model.
In Section \ref{sec:discussion}, we discuss the interpretation of these parameters obtained from our DEM simulations.

Additionally, we determine the optimal parameter set ${( \sigma_{\rm plastic}, E^{*} )}$ for $\phi_{\rm agg} = 0.3$:
\begin{equation}
{\left( \sigma_{\rm plastic}, E^{*} \right)} = {\left( 0.428~\si{MPa}, 4.96~\si{MPa} \right)}.
\end{equation}
The solid lines in Figures \ref{fig_delta_tau_Vin_phi30}(a) and \ref{fig_delta_tau_Vin_phi30}(b) represent the analytical results for this best-fit model.
The star plotted in Figure \ref{fig_Estar_sigma}(b) corresponds to the pair ${( \sigma_{\rm plastic}, E^{*} )}$ for the best fit model.

Our results highlight the necessity of elastoplastic modeling when considering the velocity dependence of $\tau_{\rm max}$ for porous aggregates.
For $\phi_{\rm agg} = 0.4$, the analytical prediction for the rigid plastic limit does not match the numerical results (Figure \ref{fig_delta_tau_Vin}(b)), whereas the analytical prediction for the perfectly elastic case cannot explain the numerical results for $\phi_{\rm agg} = 0.3$ (Figure \ref{fig_delta_tau_Vin_phi30}(b)).

\section{Discussion}
\label{sec:discussion}

In this section, we provide brief explorations of the elastic constants (Section \ref{sec:K_nu}) and yield stress of aggregates (Section \ref{sec:sigma}).
We also discuss the interaggregate motion following maximum compression (Section \ref{sec:return}).

\subsection{Bulk Modulus and Poisson's Ratio of Aggregates}
\label{sec:K_nu}

In Section \ref{sec:results}, we explored the mechanical properties of aggregates composed of monodisperse spherical ice particles with a radius of $r_{\rm p} = 100~\si{nm}$.
Our findings revealed that the reduced Young's modulus of these aggregates is $E^{*} = 15.7~\si{MPa}$ for a volume filling factor of $\phi_{\rm agg} = 0.4$.
Here, we present an order-of-magnitude estimate for other elastic constants, specifically the bulk modulus $K_{\rm agg}$ and the Poisson's ratio $\nu_{\rm agg}$.
These constants are the key parameters when we consider isotropic compression processes.

For simplicity, we assume that all particle--particle contacts within a compressed aggregate share the same values of $\delta_{\rm p}$ and $F_{\rm p}$ when the applied pressure is $P$.
The interparticle distance, $d_{\rm p}$, is defined as $d_{\rm p} = 2 r_{\rm p} - \delta_{\rm p}$.
At the equilibrium state where $F_{\rm p} = 0$ (and thus $P = 0$), $d_{\rm p}$ is equal to $2 r_{\rm p} - \delta_{0}$.
Thus, the deviation of $d_{\rm p}$ from the equilibrium state is $- {( \delta_{\rm p} - \delta_{0} )}$.
By definition, the bulk modulus of aggregate, $K_{\rm agg}$, is given as:
\begin{equation}
K_{\rm agg} = \frac{1}{3} {\left( \frac{\delta_{\rm p} - \delta_{0}}{2 r_{\rm p} - \delta_{0}} \right)}^{-1} P.
\end{equation}
The dependence of $P$ on $F_{\rm p}$ is given by the following equation \cite{https://doi.org/10.1002/cite.330420806, PhysRevE.109.024904}:
\begin{equation}
P = \frac{Z_{\rm agg} \phi_{\rm agg}}{4 \pi {r_{\rm p}}^{2}} {\left( \frac{2 r_{\rm p}}{2 r_{\rm p} - \delta_{0}} \right)}^{-1} F_{\rm p},
\end{equation}
where $Z_{\rm agg}$ represents the average coordination number within the aggregate.

We employ the JKR contact model \cite{1971RSPSA.324..301J} to describe the interparticle normal force.
When the compression length for two particles in contact is close to the equilibrium value (i.e., ${| \delta_{\rm p} - \delta_{0} |} \ll \delta_{0}$), the interparticle normal force is approximately given by the following linear relationship (see Figure \ref{fig_JKR}):
\begin{equation}
F_{\rm p} \approx k_{\rm JKR} {\left( \delta_{\rm p} - \delta_{0} \right)},
\end{equation}
where $k_{\rm JKR} = {( 6 F_{\rm c} )} / {( 5 \delta_{0} )}$ represents the spring constant.

When dust aggregates are prepared using the CPE procedure, $Z_{\rm agg}$ is a simple function of $\phi_{\rm agg}$ \cite{2011ApJ...737...36W}.
The average coordination number for CPE aggregates, $Z_{\rm CPE}$, is given by
\begin{equation}
Z_{\rm CPE} = 12 \frac{\phi_{\rm agg}}{\phi_{\rm fcc}},
\label{eq:ZCPE}
\end{equation}
where
\begin{equation}
\phi_{\rm fcc} = \frac{\pi}{3 \sqrt{2}} {\left( \frac{2 r_{\rm p}}{2 r_{\rm p} - \delta_{0}} \right)}^{3}
\end{equation}
is the maximum filling factor for the face-centered cubic lattice.
With this, we can roughly estimate $K_{\rm agg}$ as follows:
\begin{eqnarray}
K_{\rm agg} & \approx & \frac{36 \sqrt{2}}{5 \pi^{2}} {\left( \frac{2 r_{\rm p}}{2 r_{\rm p} - \delta_{0}} \right)}^{-5} \frac{F_{\rm c} {\phi_{\rm agg}}^{2}}{\delta_{0} r_{\rm p}} \nonumber \\
            & \approx & 74.3 {\left( \frac{\phi_{\rm agg}}{0.4} \right)}^{2}~\si{MPa}.
\end{eqnarray}

We can also estimate the value of $\nu_{\rm agg}$ using $K_{\rm agg}$ and $E^{*}$.
The relationship between $K_{\rm agg}$ and $E^{*}$ is given by
\begin{equation}
E^{*} = \frac{3 {\left( 1 - 2 \nu_{\rm agg} \right)}}{2 {\left( 1 - {\nu_{\rm agg}}^{2} \right)}} K_{\rm agg},
\end{equation}
where $\nu_{\rm agg}$ is the Poisson's ratio of aggregates.
With the given values of $E^{*} = 15.7~\si{MPa}$ and $K_{\rm agg} = 74.3~\si{MPa}$ for $\phi_{\rm agg} = 0.4$, we find that $\nu_{\rm agg}$ is approximately
\begin{equation}
\nu_{\rm agg} = 0.443.
\end{equation}
It's worth noting that the lower and upper limits for stable isotropic materials are $-1$ and $0.5$ \cite{2011NatMa..10..823G}, and our estimate of $\nu_{\rm agg} = 0.443$ falls within this range.
However, our estimation of $K_{\rm agg}$ may not be precise, potentially containing a significant error.
We have simplified our model by neglecting variations in $\delta_{\rm p}$ and $F_{\rm p}$ within a compressed aggregate, while in reality, such variations exist \cite{1998PhRvE..57.3164M, 2001PhRvL..86..111O}.
Further investigations into the elastic properties of dust aggregates are necessary.

\subsection{Yield Stress of Aggregates}
\label{sec:sigma}

In Section \ref{sec:velocity_theoretical}, we reported yield stress of $\sigma_{\rm plastic} = 2.25~\si{MPa}$ and $0.428~\si{MPa}$ for CPE aggregates with $\phi_{\rm agg} = 0.4$ and $0.3$ in our DEM simulations.
In this section, we demonstrate that the $\sigma_{\rm plastic}$ value obtained from our collision simulations is notably larger than that estimated from quasistatic compression simulations \cite{2023ApJ...953....6T}.
Additionally, we discuss potential explanations for the significant discrepancy in the yield stress.

\begin{figure}
\centering
\includegraphics[width = \columnwidth]{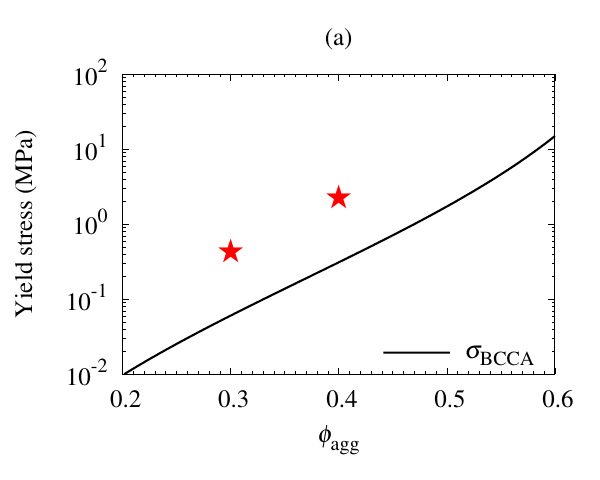}
\includegraphics[width = \columnwidth]{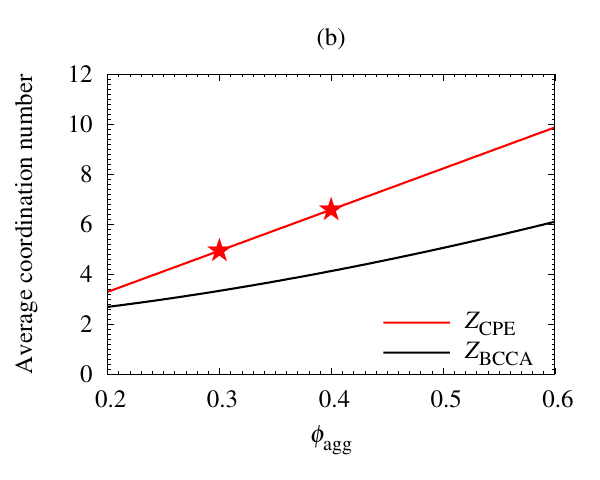}
\caption{
Yield stress and average coordination number against $\phi_{\rm agg}$.
(a) Yield stress of compressed BCCA aggregates, $\sigma_{\rm BCCA}$ (black line, Equation (\ref{eq:sigma_BCCA})).
The red stars represent the yield stress of CPE aggregates with $\phi_{\rm agg} = 0.4$ ($\sigma_{\rm plastic} = 2.25~\si{MPa}$) and $\phi_{\rm agg} = 0.3$ ($\sigma_{\rm plastic} = 0.428~\si{MPa}$).
(b) Average coordination number.
The black line represents the filling factor dependence for compressed BCCA aggregates, $Z_{\rm BCCA}$ (Equation (\ref{eq:Z_BCCA})), whereas the red line is that for CPE aggregates, $Z_{\rm CPE}$ (Equation (\ref{eq:ZCPE})).
}
\label{fig_phi_sigma_Z}
\end{figure}

Tatsuuma et al.~\cite{2023ApJ...953....6T} conducted numerical simulations of the quasistatic compression of dust aggregates consisting of monodisperse spherical ice particles with a radius of $r_{\rm p} = 100~\si{nm}$.
They used highly porous aggregates prepared by ballistic cluster--cluster aggregation (BCCA) process as initial conditions.
In their simulations, aggregates underwent isotropic compression facilitated by moving periodic boundaries (see their Figure 1).
Figure \ref{fig_phi_sigma_Z}(a) shows the yield stress with respect to $\phi_{\rm agg}$.
They derived an empirical formula for the yield stress, $\sigma_{\rm BCCA}$ (black solid line):
\begin{equation}
\sigma_{\rm BCCA} = 0.47 {\left( \frac{1}{\phi_{\rm agg}} - \frac{1}{\phi_{\rm max}} \right)}^{- 3.0}~\si{MPa},
\label{eq:sigma_BCCA}
\end{equation}
where $\phi_{\rm max} = 0.74$.
It is noteworthy that we observed $\sigma_{\rm plastic}$ for CPE aggregates (red stars) to be several times larger than $\sigma_{\rm BCCA}$. 

One plausible explanation for this substantial discrepancy could be the considerable difference in the average coordination number.
The rearrangement of constituent particles associated with interparticle tangential motions occurs during the compression of dust aggregates \cite{2017P&SS..149...14O}.
Aggregates with higher average coordination numbers would have larger yield stress.\footnote{
To the best of our knowledge, the dependence of the yield stress on the average coordination number has never been investigated, however.}
The average coordination number is known to be strongly influenced by the preparation procedure of aggregates.
In the case of compressed BCCA aggregates (black line), Arakawa et al.~\cite{2019Icar..324....8A, 2019PTEP.2019i3E02A} found that
\begin{equation}
Z_{\rm BCCA} = 2 + 9.38 {\phi_{\rm agg}}^{1.62},
\label{eq:Z_BCCA}
\end{equation}
while CPE aggregates (red line) exhibit a dependence on the filling factor as given by Equation (\ref{eq:ZCPE}).
As demonstrated in Figure \ref{fig_phi_sigma_Z}(b), CPE aggregates exhibit a higher average coordination number compared to compressed BCCA aggregates.
This difference in average coordination number could be a key factor in why CPE aggregates exhibit higher yield stress than compressed BCCA aggregates.

Another possible explanation is that the yield stress value obtained from collision simulations may inherently be higher than that derived from static compression simulations.
Tanaka et al.~\cite{2023ApJ...945...68T} investigated the compression of porous aggregates through sticking collisions with high mass ratios, employing sequential dust collisions.
They utilized ice grains with a radius of $r_{\rm p} = 100~\si{nm}$ as constituent particles, and the collision velocity was set to $v_{\rm col} = 1.8$--$7.1~\si{m.s^{-1}}$.
Their findings revealed that the energy fraction contributing to the work for compression is approximately 40\%.
This suggests that the apparent $\sigma_{\rm plastic}$ value for collisional compression is two and a half times larger than that for quasistatic compression.

The discrepancy between the yield stress values obtained from collisional and static compression tests has been observed not only in numerical simulations but also in laboratory experiments.
Katsuragi and Blum \cite{2017ApJ...851...23K} conducted collisional compression experiments by dropping a solid sphere onto a dust aggregate.
They investigated the motion of the impinging projectile using a high-speed camera, and the yield stress of the aggregate was determined from the penetration motion. 
The aggregate used in their study consisted of monodisperse spherical ${\rm Si}{\rm O}_{2}$ grains with a radius of $r_{\rm p} = 0.76~\si{\micro m}$, and the filling factor of aggregate was $\phi_{\rm agg} = 0.35$.
They found that the motion of the impinging projectile is well explained when the yield stress is $0.12~\si{MPa}$; however, this value is approximately one order of magnitude higher than that reported in static compression tests ($1.3 \times 10^{-2}~\si{MPa}$ \cite{2009ApJ...701..130G}).
Further investigations into the cause of large $\sigma_{\rm plastic}$ value are essential.

\subsection{Return Phase}
\label{sec:return}

We have demonstrated that the compressive behavior of dust aggregates aligns well with that of elastoplastic spheres.
However, it is crucial to investigate the interaggregate motion in the return phase (i.e., the phase where the mutual velocity $v$ is negative; see Appendix \ref{app:Andrews}) to gain a comprehensive understanding of the final outcomes upon collision.

Arakawa et al.~\cite{2023ApJ...951L..16A} reported that the outcomes of aggregate--aggregate collisions are stochastic, and the sticking probability decreases with increasing the size of aggregates.
However, these findings contradict the prediction of Andrews' model \cite{doi:10.1080/14786443008565033}.
The model predicts that collisions between two elastoplastic spheres always result in bouncing, and the coefficient of restitution is independent of the size of aggregates.

\begin{figure}
\centering
\includegraphics[width = \columnwidth]{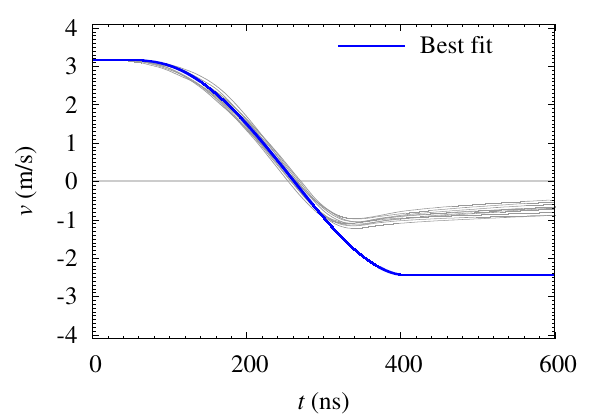}
\caption{
Temporal evolution of the mutual velocity for $\phi_{\rm agg} = 0.4$, $R_{\rm agg} = 80 r_{\rm p}$, and $v_{\rm col} = 3.2~\si{m.s^{-1}}$ (see also Figure \ref{fig_temporal}(b)).
Gray lines represent the numerical results for 10 runs.
The blue line is the analytical result based on Andrews' model with the best-fit parameters ($\sigma_{\rm plastic} = 2.25~\si{MPa}$ and $E^{*} = 15.7~\si{MPa}$).
}
\label{fig_return}
\end{figure}

In this context, we find that Andrews' model \cite{doi:10.1080/14786443008565033} for elastoplastic spheres fails to reproduce the numerical results of the interaggregate motion during the return phase.
Figure \ref{fig_return} illustrates the temporal evolution of the mutual velocity for $\phi_{\rm agg} = 0.4$, $R_{\rm agg} = 80 r_{\rm p}$, and $v_{\rm col} = 3.2~\si{m.s^{-1}}$, which are consistent with those used in Figure \ref{fig_temporal}(b).
The gray lines represent the numerical results for 10 runs, while the blue line depicts the analytical result based on Andrews' model using the best-fit parameters (i.e., $\sigma_{\rm plastic} = 2.25~\si{MPa}$ and $E^{*} = 15.7~\si{MPa}$; see Figure \ref{fig_Estar_sigma}(a)). 
The theoretical model successfully reproduces the temporal evolution of $v$ for compressive motion with $v > 0$.
However, numerical results deviate from the theoretical model in the return phase with $v < 0$.

We also observe that the return motion decelerates (i.e., $- v$ starts to decrease) in the middle of the return phase in simulations.
Interestingly, this contradicts the theoretical prediction of Andrews \cite{doi:10.1080/14786443008565033}.
In the theoretical model, the interaggregate force $F$ is always positive (i.e., repulsive).
However, in numerical simulations, the interaggregate force is initially repulsive but subsequently turns into an attractive force.
This fact implies that the tensile interaction near the interaggregate contact area, which is not considered in Andrews' model, plays a crucial role in the return phase.

\begin{figure}
\centering
\includegraphics[width = \columnwidth]{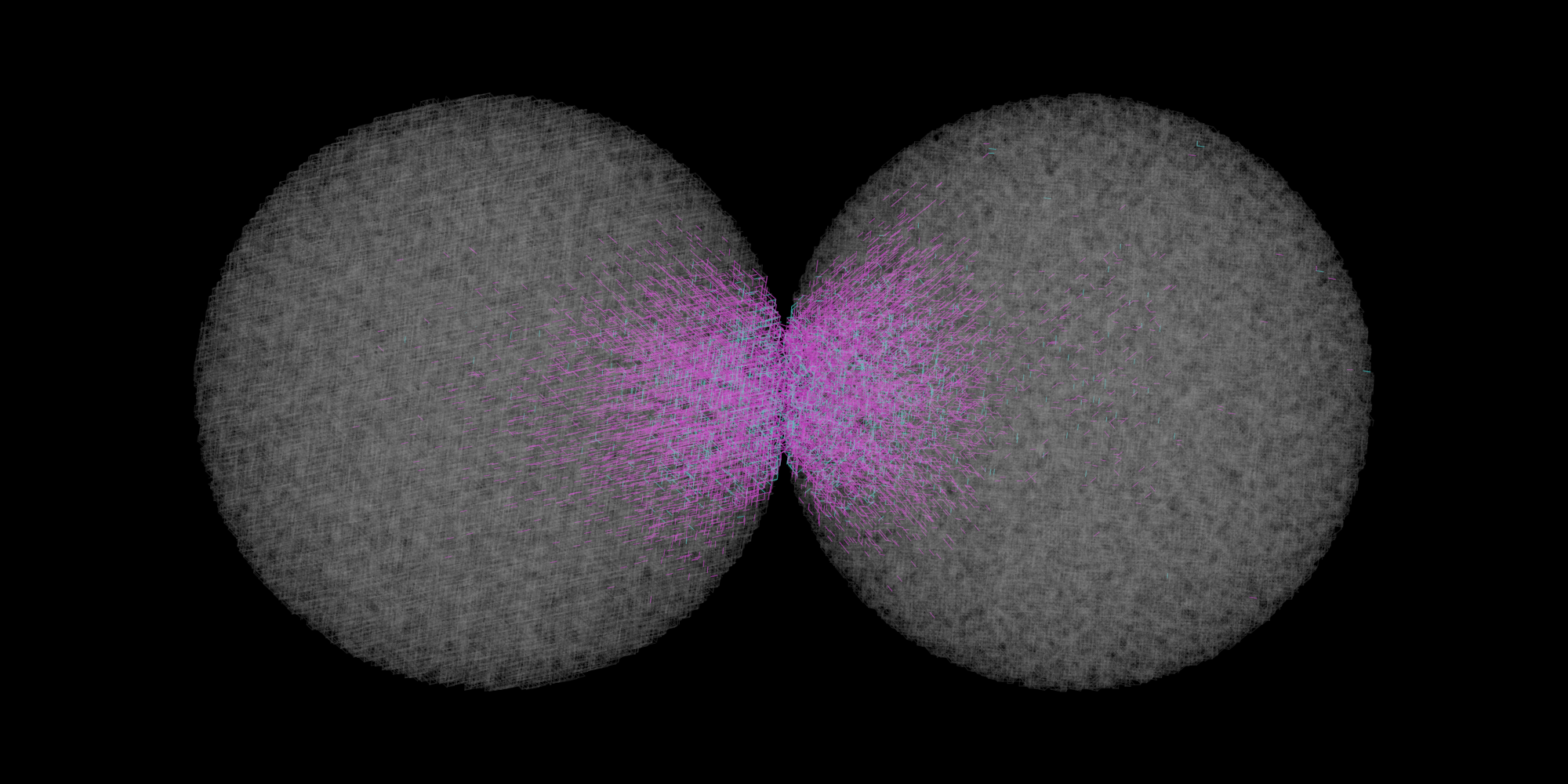}
\includegraphics[width = \columnwidth]{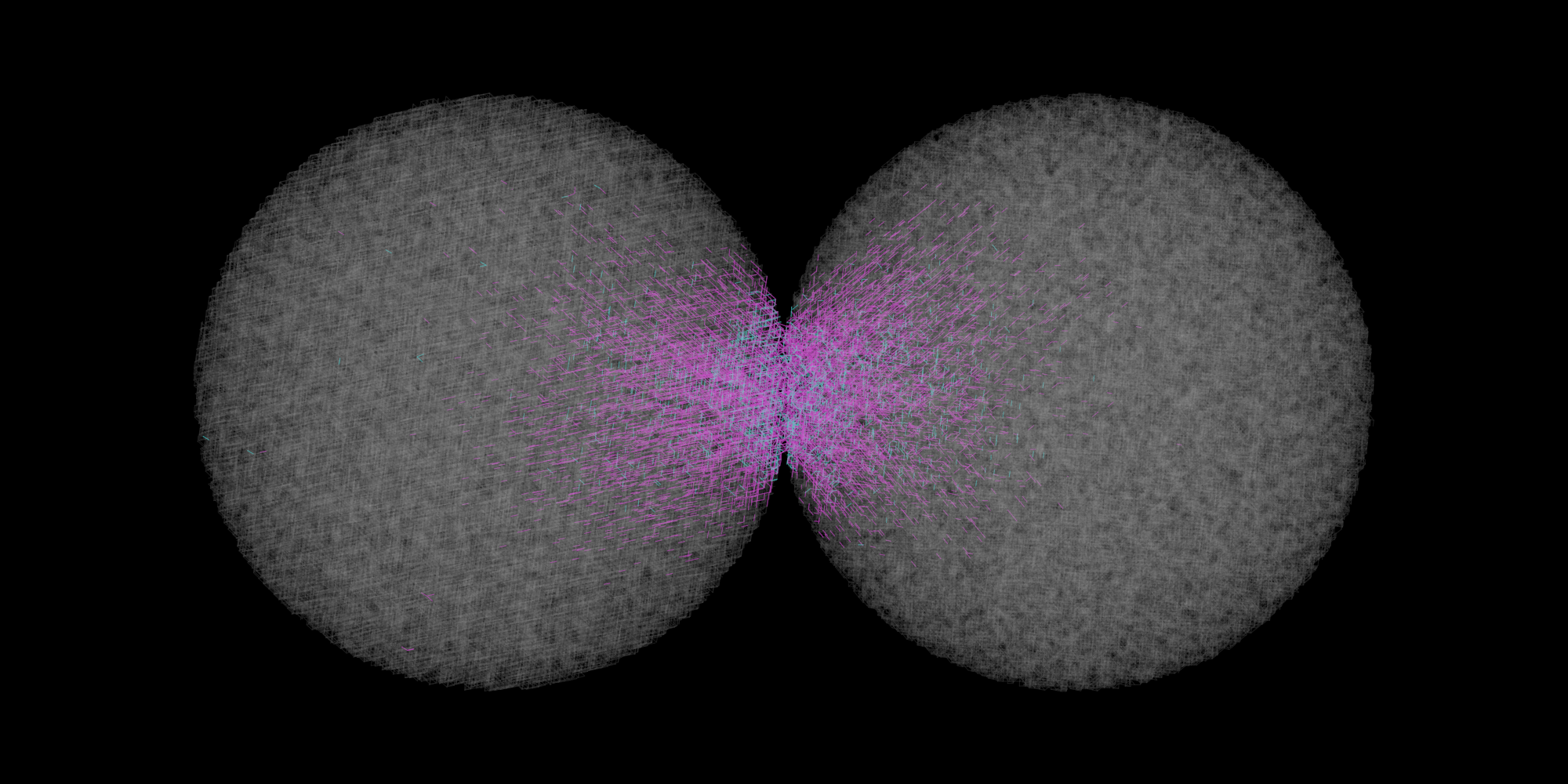}
\includegraphics[width = \columnwidth]{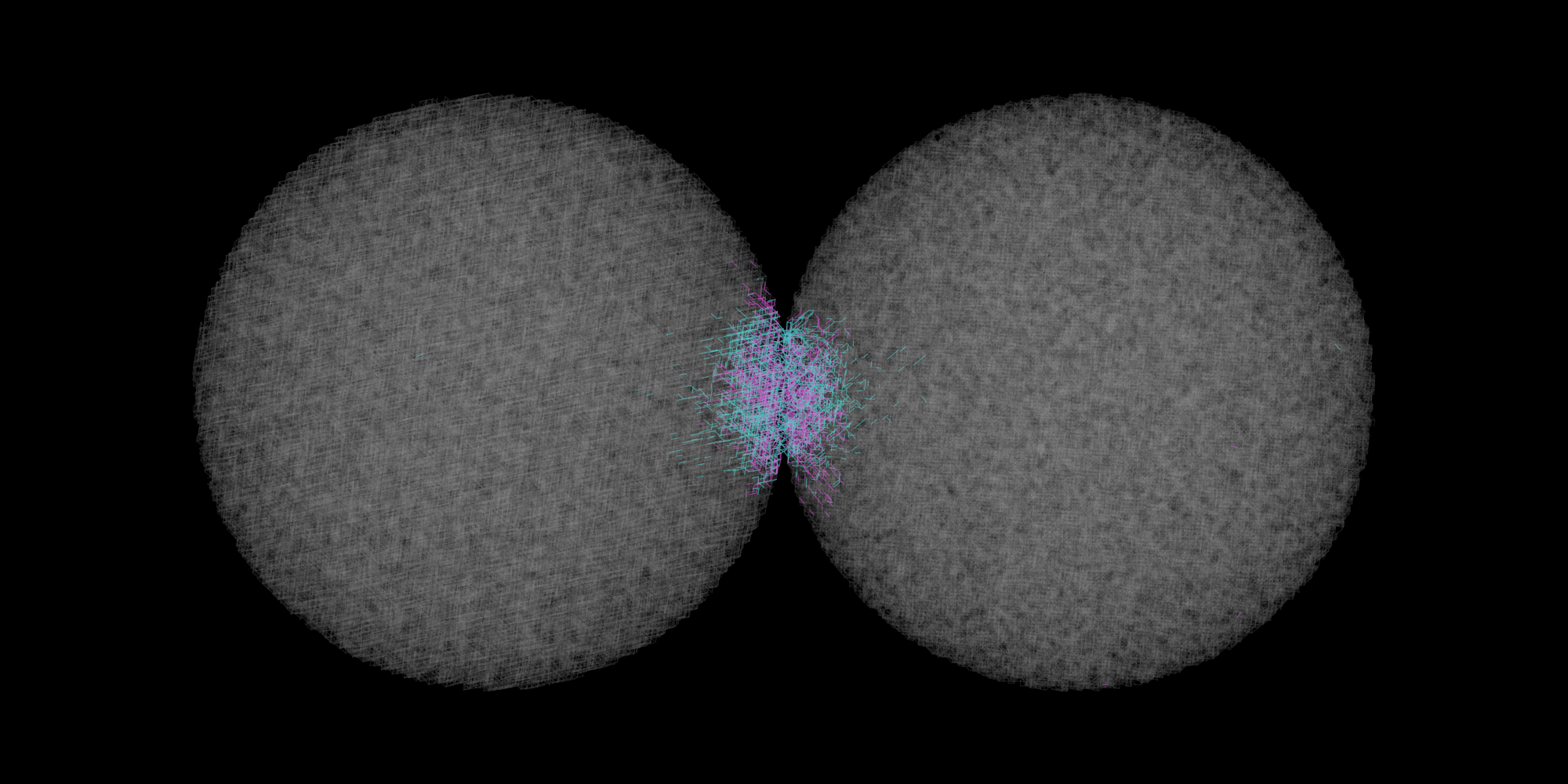}
\includegraphics[width = \columnwidth]{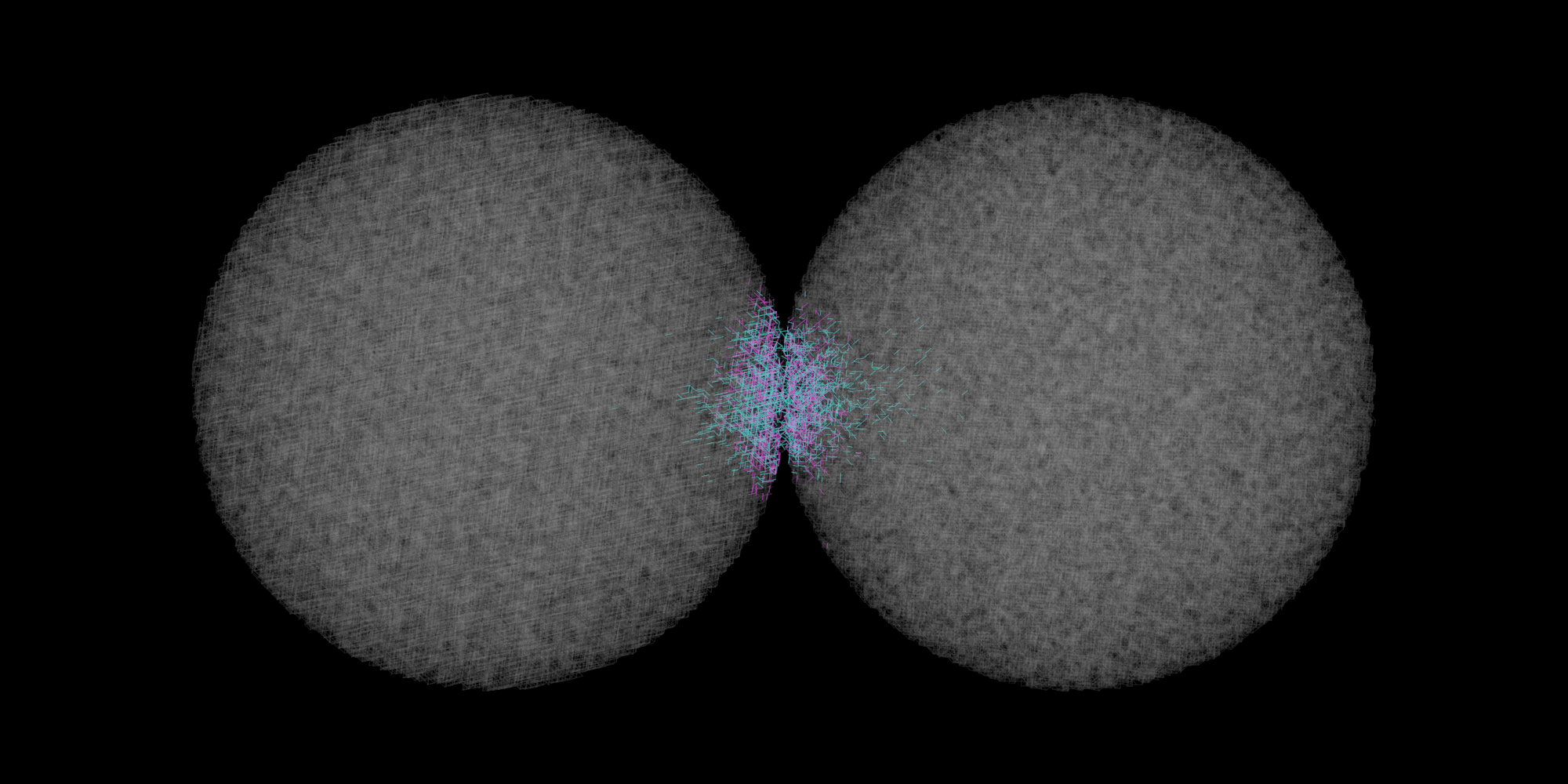}
\caption{
Distribution of the interparticle normal forces and their temporal evolution.
Here, we illustrate the result for the run shown in Figure \ref{fig_snapshot}.
Magenta, cyan, and white lines represent particle--particle contacts with a compressive force of $F_{\rm p} > 0.2 F_{\rm c}$, contacts with a tensile force of $F_{\rm p} < - 0.2 F_{\rm c}$, and others, respectively.
The first snapshot is taken at $t = 250~\si{ns}$, and the time interval is $50~\si{ns}$.
}
\label{fig_snapshot_force}
\end{figure}

Figure \ref{fig_snapshot_force} illustrates the distribution of interparticle normal forces and their temporal evolution in the simulation run shown in Figure \ref{fig_snapshot}.
Magenta, cyan, and white lines represent particle--particle contacts with a compressive force of $F_{\rm p} > 0.2 F_{\rm c}$, contacts with a tensile force of $F_{\rm p} < - 0.2 F_{\rm c}$, and others, respectively.
In this simulation run, $v$ reaches its minimum at $t \approx 344~\si{ns}$.
The first and second snapshots are taken at $t = 250~\si{ns}$ and $300~\si{ns}$, revealing that the compressive interaction dominates the tensile interaction for particle--particle contacts near the interaggregate contact area.
Conversely, the third and fourth snapshots taken at $t = 350~\si{ns}$ and $400~\si{ns}$ highlight the significant role played by the tensile interaction.
The transition from compressive to tensile interaction approximately corresponds to the time when $- v$ reaches its maximum.
The energy dissipation in the return phase is the key to understanding the size- and velocity-dependent sticking probability of dust aggregates, and further investigations are needed in future studies.

\section{Conclusion}

Aggregates consisting of micron- and submicron-sized cohesive dust grains are ubiquitous in nature, and understanding their collisional behavior is essential.
It is known that low-speed collisions of dust aggregates result in either sticking or bouncing.
Recently, Arakawa et al.~\cite{2023ApJ...951L..16A} revealed that the sticking probability decreases with increasing the size of aggregates; however, the reason for this dependence is still unclear.
As bouncing collisions among porous aggregates result in significant energy loss associated with local and permanent compaction, we expect that exploring the plastic deformation of aggregates during collisional compression provides us with the key to understanding their behavior.

In this study, we conducted DEM simulations of collisions between two aggregates of submicron-sized ice particles and investigated their compressive behavior.
We found that the size- and velocity-dependence of the maximum compression length is in excellent agreement with that predicted from Andrews' model \cite{doi:10.1080/14786443008565033} of elastoplastic spheres (Section \ref{sec:results}).
Additionally, we derived the best-fit parameters of the yield stress and the reduced Young's modulus of aggregates: $\sigma_{\rm plastic} = 2.25~\si{MPa}$ and $E^{*} = 15.7~\si{MPa}$ for $\phi_{\rm agg} = 0.4$, and $\sigma_{\rm plastic} = 0.428~\si{MPa}$ and $E^{*} = 4.96~\si{MPa}$ for $\phi_{\rm agg} = 0.3$.
The plastic deformation during collisional compression is more significant for $\phi_{\rm agg} = 0.3$ than for $\phi_{\rm agg} = 0.4$.
It should be noted that the $\sigma_{\rm plastic}$ value obtained from our collision simulations is notably larger than that acquired from quasistatic compression simulations \cite{2023ApJ...953....6T}.
We concisely remarked on the plausible causes of this discrepancy in Section \ref{sec:sigma}.

In contrast, we revealed that the interaggregate motion during the return phase cannot be explained by Andrews' model \cite{doi:10.1080/14786443008565033} (Section \ref{sec:return}).
The outcomes of aggregate--aggregate collisions are stochastic, and the interaggregate force is initially repulsive but subsequently turns into an attractive force.
These findings contradict Andrews' model \cite{doi:10.1080/14786443008565033}.
We speculate that the size dependence of the sticking probability reported in previous studies \cite{2023ApJ...951L..16A} would predominantly originate from the energy dissipation process in the return phase and its size dependence.
We will test this hypothesis in future studies.

We acknowledge that the size and velocity ranges investigated in this study are still limited, and further investigations are necessary in future studies.
In this study, the filling factor of aggregates is fixed at $\phi_{\rm agg} = 0.4$ or $0.3$, and we will investigate the filling factor dependence in a forthcoming study.
In addition, the parameter calibration by comparing numerical results with the ground truth acquired from laboratory experiments would be helpful.
We focused on aggregates consisting of spherical particles; however, many natural aggregates, including snowflakes and graupel particles, consist of irregular-shaped grains \cite{2019JGRD..12410049L}.
Impacts of particle shapes on the collisional behavior should be investigated in future studies using both DEM simulations and laboratory experiments.

\backmatter


\bmhead{Acknowledgments}

S.A.~was supported by JSPS KAKENHI Grant No.~JP24K17118.
Numerical computations were carried out on the PC cluster at CfCA, NAOJ.
We thank Dr.~Yukari M.~Toyoda for fruitful discussions and comments.

\section*{Declarations}

\bmhead{Conflict of interest}

The authors have no conflicts of interest directly relevant to the content of this article.

\bmhead{Data availability}

The data that support the findings of this study are available from the corresponding author upon request.


\clearpage

\begin{appendices}

\section{Interparticle Tangential Motion}
\label{app:tangential}

\begin{figure*}
\centering
\includegraphics[width = 0.8\textwidth]{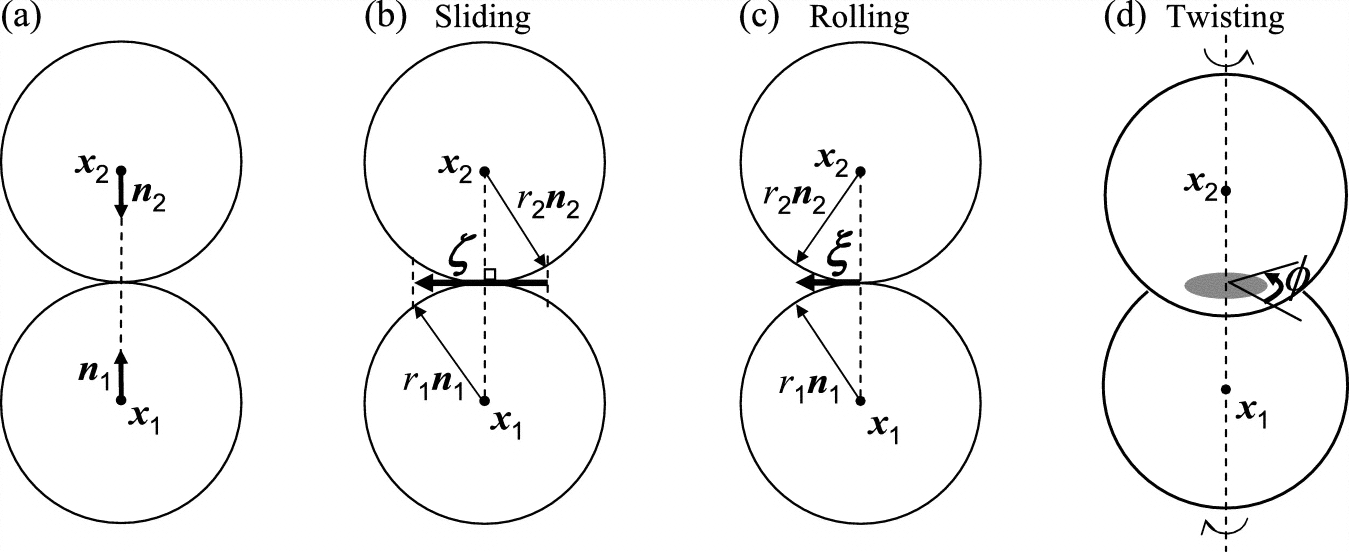}
\caption{
Schematics of (a) the beginning of contact, (b) sliding, (c) rolling, and (d) twisting displacements.
The contact pointers for the particles $1$ and $2$ are represented by $\bm{n}_{1}$ and $\bm{n}_{2}$, respectively.
In this study, aggregates consisting of monodisperse spherical particles are considered, and the radii of the particles $1$ and $2$, $r_{1}$ and $r_{2}$, are set to $r_{1} = r_{2} = r_{\rm p}$.
Note that the displacements are exaggerated compared to the actual ones.
Figure taken from Wada et al.~\cite{2007ApJ...661..320W}.
Reproduced with permission from the American Astronomical Society.
}
\label{fig_Wada07}
\end{figure*}

The tangential motion between two particles in contact can be divided into three types: sliding, rolling, and twisting.
The displacements corresponding to these tangential motions are represented by the rotation of the two particles.
Here, we elaborate on these displacements and the resistances against them (Figure \ref{fig_Wada07}).

The tangential interaction model employed in this study was developed by Wada et al.~\cite{2007ApJ...661..320W}.
We treat tangential interactions as linear elastic springs when all displacements remain below specified thresholds.
Additionally, we account for inelastic behavior when certain displacements surpass these threshold values.

We use contact pointers, represented as unit vectors originating from the center of each particle and directed toward the contact point \cite{2002Icar..157..173D}.
At the initial contact moment ($t = t_{0}$), the contact pointer of the particle $1$, $\bm{n}_{1}$, is set to $\bm{n}_{1} = - \bm{n}_{\rm c}$.
The sliding and rolling displacements, $\bm{\zeta}$ and $\bm{\xi}$, are defined by
\begin{equation}
\bm{\zeta} = r_{\rm p} {\left[ \bm{n}_{1} - \bm{n}_{2} - {\left( \bm{n}_{1} \cdot \bm{n}_{\rm c} - \bm{n}_{2} \cdot \bm{n}_{\rm c} \right)} \bm{n}_{\rm c} \right]},
\end{equation}
and
\begin{equation}
\bm{\xi} = r_{\rm p} \frac{\bm{n}_{1} + \bm{n}_{2}}{2}.
\end{equation}
The twisting displacement, $\bm{\phi}$, is given by 
\begin{equation}
\bm{\phi} = \phi \bm{n}_{\rm c},
\end{equation}
where $\phi$ represents the twisting angle.
At $t = t_{0}$, all tangential displacements are initialized to $\bm{0}$, and inelastic motions commence when certain displacements exceed critical values, namely, $\zeta_{\rm crit}$, $\xi_{\rm crit}$, and $\phi_{\rm crit}$.

The twisting angle, $\phi$, is given by the following equation:
\begin{equation}
\phi {( t )} = \int_{t_{0}}^{t} \mathscr{I} {( t^{'} )} {\left[ \bm{\omega}_{1} {( t^{'} )} - \bm{\omega}_{2} {( t^{'} )} \right]} \cdot \bm{n}_{\rm c} {( t^{'} )}~{\rm d}{t^{'}},
\end{equation}
where $\bm{\omega}_{i}$ represents the angular velocity of the $i$th particle, and $\mathscr{I}$ denotes the switch between elastic and inelastic interactions.
We define $\mathscr{I} = 0$ when $\phi = \pm \phi_{\rm crit}$ and $\phi {\left[ {\left( \bm{\omega}_{1} - \bm{\omega}_{2} \right)} \cdot \bm{n}_{\rm c} \right]} > 0$, indicating inelastic behavior.
Otherwise, $\mathscr{I} = 1$, representing elastic interactions.

To compute $\bm{\zeta}$ and $\bm{\xi}$ at each time step, we track the contact pointers $\bm{n}_{1}$ and $\bm{n}_{2}$.
The rotational transformation of the contact pointers is expressed using a rotation matrix.
Further details are described in Section 2.2 of Wada et al.~\cite{2007ApJ...661..320W}.

The torques on the particle 1 exerted by the particle 2 are given as follows:
\begin{equation}
\bm{M}_{{\rm s}, 12} = - r_{\rm p} k_{\rm s} \bm{n}_{1} \times \bm{\zeta},
\end{equation}
\begin{equation}
\bm{M}_{{\rm r}, 12} = - \frac{r_{\rm p}}{2} k_{\rm r} \bm{n}_{1} \times \bm{\xi},
\end{equation}
and
\begin{equation}
\bm{M}_{{\rm t}, 12} = - k_{\rm t} \bm{\phi}.
\end{equation}
Here, $k_{\rm s}$, $k_{\rm r}$, and $k_{\rm t}$ are the spring constants for each interaction.
The force on the particle 1 due to sliding motion with the particle 2 is given by
\begin{equation}
\bm{F}_{{\rm s}, 12} = - k_{\rm s} \bm{\zeta} \frac{r_{\rm p} {\left( \bm{n}_{2} - \bm{n}_{1} \right)} \cdot \bm{n}_{\rm c}}{|| \bm{x}_{1} - \bm{x}_{2} ||}.
\end{equation}

Both the critical displacement and the spring constant for each interaction are equal to those of previous studies \cite{2011ApJ...737...36W, 2023ApJ...951L..16A}.
The critical displacements for sliding, rolling, and twisting are given by
\begin{equation}
\zeta_{\rm crit} = \frac{2 - \nu_{\rm p}}{16 \pi} a_{0},
\end{equation}
\begin{equation}
\xi_{\rm crit} = 0.8~\si{nm},
\end{equation}
and
\begin{equation}
\phi_{\rm crit} = \frac{1}{16 \pi},
\end{equation}
respectively.
The spring constants are given by
\begin{eqnarray}
k_{\rm s} & = & \frac{2 a_{0} E_{\rm p}}{{( 2 - \nu_{\rm p} )} {( 1 + \nu_{\rm p} )}}, \nonumber \\
k_{\rm r} & = & \frac{8 F_{\rm c}}{r_{\rm p}}, \nonumber \\
k_{\rm t} & = & \frac{4 E_{\rm p} {a_{0}}^{3}}{3 {( 1 + \nu_{\rm p} )}}.
\end{eqnarray}

\section{Contact Mechanics for Elastoplastic Spheres}
\label{app:Andrews}

\begin{figure}
\centering
\includegraphics[width = 0.6\columnwidth]{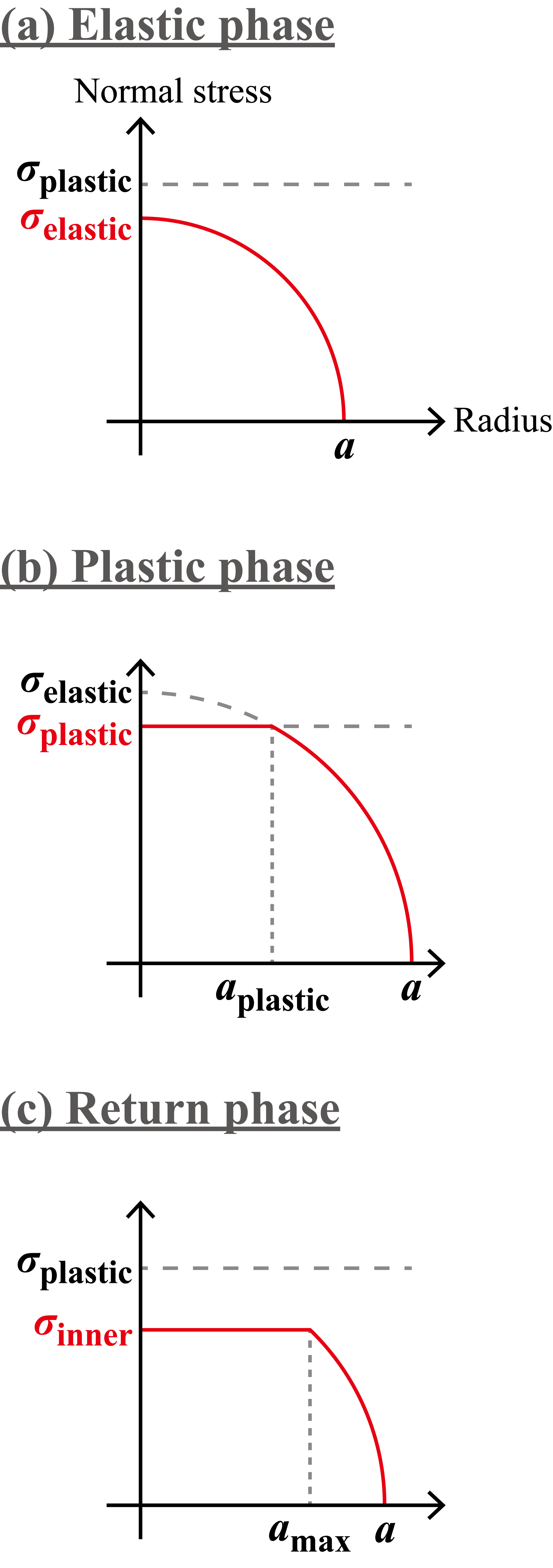}
\caption{
Schematic of the spatial distribution of the normal stress within the contact area.
(a) Elastic phase: the normal stress at the center of the contact area is below the threshold for plastic deformation, $\sigma_{\rm plastic}$.
(b) Plastic phase: the normal stress at the center of the contact area reaches the threshold, and the compaction proceeds.
(c) Return phase following the plastic phase.
}
\label{fig_pressure}
\end{figure}

The compressive behavior of spherical dust aggregates during low-speed head-on collisions is well-approximated by that of elastoplastic spheres.
Here, we introduce a contact model for elastoplastic spheres, originally derived by Andrews \cite{doi:10.1080/14786443008565033} (Figure \ref{fig_pressure}).

We consider head-on collisions of equal-size aggregates with an effective radius of $R_{\rm agg, eff}$.
The spatial distribution of the normal stress ($\sigma$) within the interaggregate contact area is a function of the compression length, $\delta$, and the distance from the center of the contact area, $a^{'}$: $\sigma = \sigma {( \delta, a^{'} )}$.
The interaggregate repulsive force, $F$, depends on $\delta$:
\begin{equation}
F {( \delta )} = \int_{0}^{a}~2 \pi a^{'} \sigma {( \delta, a^{'} )}~{\rm d}a^{'},
\label{eq:F_definition}
\end{equation}
where $a = \sqrt{R^{*} \delta}$ represents the interaggregate contact radius \cite{hertz1896miscellaneous}.

\subsection{Hertzian Model for Elastic Spheres}

When the normal stress at the center of the contact area is below the threshold for plastic deformation, the spatial distribution of the normal stress is modeled by the Hertzian model for perfectly elastic spheres \cite{hertz1896miscellaneous} (see Figure \ref{fig_pressure}(a)):
\begin{eqnarray}
\sigma {( \delta, a^{'} )} & = & \sigma_{\rm elastic} {( \delta, a^{'} )} \nonumber \\
                        & = & \frac{2 E^{*} \sqrt{\delta}}{\pi \sqrt{R^{*}}} \sqrt{ 1 - \frac{{a^{'}}^{2}}{R^{*} \delta} },
\label{eq:sigma_dist_Hertz}
\end{eqnarray}
where $E^{*}$ is the reduced Young's modulus of aggregates.
By solving Equations (\ref{eq:F_definition}) and (\ref{eq:sigma_dist_Hertz}), the repulsive force is then given by 
\begin{equation}
F {( \delta )} = \frac{4 E^{*} \sqrt{R^{*}} \delta^{3/2}}{3}.
\label{eq:F_elastic_app}
\end{equation}

\subsection{Andrews' model for Elastoplastic Spheres}

The Hertzian model for perfectly elastic spheres is not applicable when the normal stress at the center of the contact area reaches the threshold for plastic deformation.
Andrews \cite{doi:10.1080/14786443008565033} proposed an approximate model applicable to the elastoplastic deformation of spheres.

\subsubsection{Plastic Phase}

We assume that dust aggregates behave as elastic--perfectly plastic materials (Figure \ref{fig_pressure}(b)):
\begin{equation}
\sigma {( \delta, a^{'} )} = \min{\left[ \sigma_{\rm elastic} {( \delta, a^{'} )}, \sigma_{\rm plastic} \right]},
\label{eq:sigma_dist_plastic}
\end{equation}
where $\sigma_{\rm plastic}$ is the yield stress.
The critical compression length for yielding, $\delta_{\rm crit}$, is given by
\begin{equation}
\delta_{\rm crit} = {\left( \frac{\pi \sigma_{\rm plastic}}{2 E^{*}} \right)}^{2} R^{*},
\label{eq:delta_crit}
\end{equation}
which is derived from Equation (\ref{eq:sigma_dist_Hertz}).
Two colliding aggregates behave as perfectly elastic spheres when $\delta < \delta_{\rm crit}$, and their elastoplastic deformation starts at $\delta_{\rm crit}$.
At $\delta > \delta_{\rm crit}$, the inner region of the contact area with a radius of $a_{\rm plastic}$ is plastically deformed.
The radius of the plastically deformed area is
\begin{equation}
a_{\rm plastic} = \sqrt{R^{*} {\left( \delta - \delta_{\rm crit} \right)}},
\label{eq:a_plastic}
\end{equation}
which is derived from Equations (\ref{eq:sigma_dist_Hertz}) and (\ref{eq:delta_crit}).
Using Equations (\ref{eq:F_definition}) and (\ref{eq:sigma_dist_plastic}), the repulsive force is given by
\begin{equation}
F {( \delta )} = \pi R^{*} \sigma_{\rm plastic} {\left( \delta - \frac{1}{3} \delta_{\rm crit} \right)}.
\label{eq:F_plastic_app}
\end{equation}
Thornton and Ning \cite{Thornton+1998} have independently derived a mathematically identical result.

\subsubsection{Return Phase}

When the compression length reaches the maximum compression length, the radius of the plastically deformed area also takes the maximum, and it is given by 
\begin{equation}
a_{\rm max} = \sqrt{R^{*} {\left( \delta_{\rm max} - \delta_{\rm crit} \right)}},
\label{eq:a_max}
\end{equation}
which is derived from Equation (\ref{eq:a_plastic}).

Here, we consider the spatial distribution of the normal stress in the return phase following the plastic phase (Figure \ref{fig_pressure}(c)).
We assume that the outer region of the contact area of $a^{'} \ge a_{\rm max}$ is intact, and $\sigma {( \delta, a^{'} )}$ is given by Equation (\ref{eq:sigma_dist_Hertz}).
In contrast, the inner region with a radius of $a_{\rm max}$ has been deformed plastically, and $\sigma {( \delta, a^{'} )}$ is given by
\begin{equation}
\sigma {( \delta, a^{'} )} = \min{\left[ \sigma_{\rm elastic} {( \delta, a^{'} )}, \sigma_{\rm inner} \right]},
\label{eq:sigma_dist_return}
\end{equation}
where $\sigma_{\rm inner}$ is the normal stress in the inner region of $a^{'} < a_{\rm max}$.
Andrews \cite{doi:10.1080/14786443008565033} simply assumed that 
\begin{equation}
\sigma_{\rm inner} {( \delta )} = \sigma_{\rm elastic} {( \delta, a_{\rm max} )}.
\end{equation}
Using Equations (\ref{eq:F_definition}) and (\ref{eq:sigma_dist_return}), the repulsive force is
\begin{eqnarray}
F {( \delta )} & = & \frac{4 E^{*} \sqrt{R^{*}} \delta^{3/2}}{3} \nonumber \\
               &   & \cdot \sqrt{ 1 - \frac{\delta_{\rm return}}{\delta} } {\left( 1 + \frac{\delta_{\rm return}}{2 \delta} \right)},
\end{eqnarray}
where $\delta_{\rm return}$ is the compression length at the breakup of the interaggregate contact.
As $a = a_{\rm max}$ at $\delta = \delta_{\rm return}$, $\delta_{\rm return}$ is derived from Equation (\ref{eq:a_max}) as:
\begin{equation}
\delta_{\rm return} = \delta_{\rm max} - \delta_{\rm crit}.
\end{equation}




\end{appendices}


\clearpage

\bibliography{sn-bibliography}

\end{document}